\documentclass[fleqn,10pt]{wlscirep}

\usepackage[utf8]{inputenc}
\usepackage[T1]{fontenc}

\usepackage{graphicx} 

\usepackage{hyperref}
\usepackage{comment}
\usepackage{physics}
\usepackage{bm}
\usepackage{algorithm}
\usepackage{algpseudocode}

\usepackage{import}
\usepackage{dsfont}
\usepackage{amsmath}
\usepackage{amssymb}
\usepackage{amsfonts}
\usepackage{tabularx}
\usepackage{mathtools}

\usepackage{color,colortbl}
\usepackage{xcolor}
\definecolor{darkgreen}{RGB}{0,170,0}
\definecolor{darkpink}{RGB}{255,150,180}

\definecolor{cristianocolor}{rgb}{.2,0.2,0.75}
\definecolor{lucapink}{rgb}{0.9,0.3,0.9}

\definecolor{applegreen}{rgb}{0.55, 0.71, 0.0}

\newcommand{\EqRef}[1]{Eq.~\eqref{#1}}
\newcommand{\FigRef}[1]{Fig.~\ref{#1}}

\usepackage[T1]{fontenc}    
\usepackage{url}            
\usepackage{booktabs}       
\usepackage{nicefrac}       
\usepackage{microtype}      
\usepackage{soul}

\usepackage{authblk}
\usepackage[normalem]{ulem}

\title{Adaptive Behavior with Stable Synapses}

\begin{abstract}
    
Behavioral changes in animals and humans, as a consequence of an error or a verbal instruction, can be extremely rapid. Improvement in behavioral performances are usually associated learning theories to synaptic plasticity. 
However, such rapid changes are not coherent with the timescales of biological synaptic plasticity, suggesting that the mechanism responsible for that could be a dynamical reconfiguration of the network involved, without changing its weights.
In the last few years, similar capabilities have been observed in transformers, foundational architecture in the field of machine learning that are widely used in applications such as natural language and image processing. Transformers are capable of in-context learning, the  ability to adapt and acquire new information dynamically within the context of the task or environment they are currently engaged in, without the need for significant changes to their underlying parameters. 
Building upon the notion of something unique within transformers enabling the emergence of this property, we claim that it could be supported by gain-modulation, feature extensively observed in biological networks, such as in pyramidal neurons thanks to input segregation and dendritic amplification.
We propose a constructive approach to induce in-context learning in an architecture composed of recurrent networks with gain modulation, demonstrating abilities inaccessible to standard networks. 
In particular, we show that, such architecture can dynamically implement standard gradient-based by encoding weight changes in the activity of another network. We argue that, while these algorithms are traditionally associated with synaptic plasticity, their reliance on non-local terms suggests that, in the brain, they can be more naturally realized at the level of neural circuits. 
We demonstrate that we can extend our approach to non-linear and temporal tasks and to reinforcement learning. We further validate our approach in a realistic robotic setting, demonstrating dynamic adaptation in a MuJoCo ant navigation task without synaptic updates, showcasing a neuromorphic control paradigm that operates through real-time network reconfiguration.
Our framework contributes to understanding the principles underlying in-context learning and adaptive behavior in both natural and artificial intelligence, and provides a measurable approach to distinguish between dynamic adaptation and synaptic plasticity, aiding experimental validation.

\end{abstract}

\author[1,3]{Cristiano Capone \thanks{Corresponding author: cristiano0capone@gmail.com}}
\author[1,2,3]{Luca Falorsi}

\affil[1]{Natl. Center for Radiation Protection and Computational Physics, Istituto Superiore di Sanità, 00161 Rome, Italy}
\affil[2]{PhD Program in Mathematics, Dept. of Mathematics, “Sapienza” University of Rome, 00185 Rome, Italy}
\affil[3]{These authors contributed equally}

\date{}

\begin{document}
\maketitle


\section*{Introduction}

\paragraph{Adaptive Behavior in Biological Agents}

The study of adaptive behavior in biological agents, including humans and animals, has been a longstanding subject of research \cite{lee2013contextual}. Observations of rapid behavioral shifts in response to environmental cues have raised questions about whether traditional mechanisms, such as synaptic plasticity, can fully account for such flexibility.

A broad range of neural and behavioral data indicates the existence of distinct systems for behavioral choice \cite{daw2005uncertainty,mcdougle2015explicit}. The conventional view posits that subcortical areas support habitual, stimulus-based responses, enabling reflex-based and rapid reactions. In contrast, higher cortical areas, such as the prefrontal cortex, are associated with reflective, goal-directed behaviour, which relies on cognitive action planning and deliberate evaluation.
Critically, goal-directed behavior allows for the flexible construction of value at each decision point through online planning procedures. This mechanism makes agents immediately sensitive to changes in outcome values, enabling them to dynamically adapt their actions to novel or changing circumstances \cite{o2017learning}. Studies in network neuroscience have identified dynamic network reconfiguration, characterized by the flexible recruitment and integration of neural circuits, as a fundamental mechanism supporting this adaptability across cognitive\cite{braun2015dynamic} and motor domains\cite{standage2023whole}.


Computational models within the predictive coding framework have recently provided insights into the neural mechanisms underlying dynamic adaptation. These models suggest that adaptation can be understood as a form of Bayesian inference, where the brain continuously infers latent contextual states to guide behavior \cite{annurev:/content/journals/10.1146/annurev-neuro-092322-100402}. A prominent example is the COIN model\cite{heald2021contextual,heald2023contextual}, which posits that adaptation arises through two complementary processes: proper learning, involving the creation and updating of memories, and apparent learning, which dynamically adjusts the expression of pre-existing memories based on inferred context. While these frameworks capture key phenomena of adaptive behavior, such as context-dependent single-trial learning and spontaneous recovery, a significant gap remains between these theoretical models and the neural substrates that implement these computations in the brain.

\paragraph{In-context learning as an emerging property in AI}

In machine learning, the capacity of a model to adapt its behavior, without weight updates, is referred to as in-context learning (ICL).
Initially, ICL was observed in architectures tailored for few-shot learning \cite{vinyals2016matching} or even zero-shot learning \cite{romera2015embarrassingly}. However, the game changed when it was observed that ICL emerges naturally in large-scale transformers \cite{brown2020language,singh2024transient}. These architectures leverage self-attention, a mechanism that dynamically changes the way past inputs are attended, in order to shape the current response. 
Their exceptional capability to adapt to contextual information allowed transformer-based architectures to achieve state-of-the-art performances in many domains, such as natural language processing and image analysis\cite{vaswani2017attention,achiam2023gpt,ramesh2021zero}.
While there are several intriguing explanations \cite{ramsauer2020hopfield,olsson2022context} for the emergence of ICL in transformers, a more formal understanding was pioneered by Von Oswald and collaborators \cite{von2023transformers}, who demonstrated that a linear self-attention layer can implement gradient descent steps within its forward pass. 
This intuition was empirically validated by analyzing trained transformers and was formally proven \cite{zhang2024trained}. 

Despite their success, transformer architectures face notable shortcomings, particularly in their memory requirements, which scale quadratically with sequence length. These requirements make those architectures largely biologically implausible.
Moreover, memory demands constrain their scalability and restrict their applicability to tasks involving long sequences, such as full-length document analysis or video understanding. 
To overcome these challenges, several efficient deep linear RNN architectures have been proposed \cite{gu2021efficiently, orvieto2023resurrecting}, demonstrating substantial performance gains over transformers, especially in long-sequence tasks. 

However, these deep learning architectures rely on several mechanisms that do not have an obvious correspondence in biological neural systems (such as the attention mechanism itself), leaving an unresolved gap in relating them to the in-context learning abilities exhibited by biological networks.
This work bridges this gap by introducing a constructive method to induce in-context learning in biologically plausible neural networks, demonstrating its applicability in a wide range of scenarios.

\paragraph{Biological support for in-context learning}

Building on this premise, our investigation explores how in-context learning, as demonstrated in deep learning architectures, might also be realized in biological recurrent neural networks.
Are there unique features in transformers that are also present in biological networks?
We propose that in-context learning can be facilitated by input segregation and dendritic amplification, two features widely observed in biological networks. We further argue that these mechanisms provide a biologically plausible foundation for implementing a process analogous to the attention mechanism in transformers.
Recent findings on dendritic computational properties \cite{poirazi2020illuminating} and on the complexity of pyramidal neurons dynamics \cite{larkum2013cellular} motivated the study of multi-compartment neuron models in the development of new biologically plausible learning rules \cite{urbanczik2014learning,guerguiev2017towards,sacramento2018dendritic,payeur2021burst}.
It has been proposed that segregation of dendritic input~\cite{guerguiev2017towards} (i.\,e., neurons receive sensory information and higher-order feedback in segregated compartments) and generation of high-frequency bursts of spikes \cite{payeur2021burst} would support backpropagation in biological neurons.
Recently, it has been suggested \cite{capone2023beyond} that this neuronal architecture naturally allows for orchestrating ``hierarchical imitation learning'', enabling the decomposition of challenging long-horizon decision-making tasks into simpler subtasks. They show a possible implementation of this in a two-level network, where the high-network produces the contextual signal for the low-network.
Here, we propose an architecture composed of gain-modulated recurrent networks that demonstrate remarkable in-context learning capabilities, which we refer to as 'dynamical adaptation'. Specifically, we illustrate that our biologically plausible architecture can dynamically adapt its behavior in response to feedback from the environment without altering its synaptic weights. 
We present results for supervised learning of temporal trajectories and reinforcement learning, involving non trivial input-output temporal relations.
This novel architecture aims to bridge the gap between biological-inspired in-context learning and the capabilities of artificial neural networks, offering a promising avenue for advancing our understanding of adaptive behavior in both natural and artificial intelligence domains.

\section*{Main}

\subsection*{Framework Formalization}

Consider a general learning scenario, where an agent is engaged in an environment $\alpha$ from a family of tasks $\mathcal{A}$. The agent receives temporal input $\vb{x}(t)$ and it has to compute a response $y(t)$, receiving feedback $f(t)$ from the environment in return. This can be an error, reward, or the target behavior.

We further assume that a low-level network extracts features $\vb{z}(t)$ relevant to the task family. The agent then adapts its response to the specific environment by learning readout parameters $\vb{w}$, which determine how to recombine these features to produce the correct response, $y(t) = Y (\vb{z}(t), \vb{w})$. Here $Y$ is a generic function that depends on the parameters $\vb{w}$.
These parameters are then adjusted at each time step in function of an internally computed error signal $e$. 
The overall learning process can then be generically rewritten as the following dynamical system:
\begin{equation}\label{eq:general-formulation} 
\begin{cases}
\tau_e \, \dot{\vb{e}} & =  E(\vb{e},\vb{y},\vb{f}), \\
\tau_y \, \vb{y} & =  Y(\vb{y},\vb{w},\vb{z}), \\
\tau_w \, \dot{\vb{w}} & =  W(\vb{w},\vb{z},\vb{e}).
\end{cases}
\end{equation}
The specific instantiation of the functions 
$W$ and $E$ depends on the learning algorithm employed and can encompass both supervised and reinforcement learning settings. Refer to the Methods section for their precise definitions.
In the classic view, the parameters $\vb{w}$ are interpreted as synaptic weights of a neuronal network, and the functions $E$ and $W$ describe synaptic plasticity mechanisms. However, this imposes stringent locality requirements, as all information must be present at the synapse.

Rather than interpreting $\vb{w}$ as synaptic weights, we view them as the dynamic activity of an auxiliary network that modulates the response $Y$ in real time based on feedback and past interactions. We refer to these dynamically adapting state variables as "virtual" weights. We refer to $\vb{w}$ as “virtual weights” because:
- Weights: they replace the physical weights, and they have the same role as the physical weights of parametrizing the map from the features $\vb{z}(t)$ to the output(s). 
- Virtual: they are technically not weights, since the are encoded in the activity of the readout neurons of the network W, and their activity is provided as an input current to the Y network. Using this strategy we replace the computation done by the synapses (including learning) with operations performed by the neural activity.
Dynamic adaptation offers key advantages over synaptic plasticity: 
\begin{itemize}
    \item {\bf Fast Timescales}: It naturally operates on faster timescales, making it well-suited for rapid behavioral changes
    \item{\bf Non-Locality}: Dynamic adaptation is not restricted by the stringent locality requirements of synaptic plasticity, allowing for richer forms of interaction and computation at the network level.
\end{itemize}

In the Supplementary Materials, we show that a non-local weight update rule, implemented through dynamic adaptation, enables rule switching with a single error feedback in a context-dependent classification task, whereas local synaptic plasticity drops to chance level after one error—independently of the timescales of the learning process (\FigRef{fig0}).
In the dynamic adaptation setting, the functions $E$ and $W$ in \eqref{eq:general-formulation} arise from the dynamics of the auxiliary network of virtual weights. We show that such networks can dynamically implement standard gradient-based algorithms widely used in machine learning. While these algorithms are traditionally associated with synaptic plasticity, their reliance on non-local terms suggests that, in the brain, they can be more naturally realized at the circuit level. We demonstrate that such circuits capable of dynamic learning can simply arise by tuning the static readout weights of random RNNs with multiplicative interactions, requiring only a limited number of training trajectories from the original learning algorithm applied to a subset of the task family.
The construction of our networks follows two distinct phases: a developmental or evolutionary phase, and a subsequent phase where the network leverages this structure for dynamical learning.
\begin{itemize}
    \item {\bf Phase 1: Development of the network structure.}
    In this phase, using an Algorithmic Distillation protocol\cite{laskin2022context}, we collect learning trajectories of the form $\{(\vb{z}_\alpha(t),\vb{w}_\alpha(t), \vb {e}_\alpha(t), \vb{f}_\alpha(t))_{t=0}^T\| \alpha \in \mathcal{A}_{ID}\}$ by running the learning dynamics in \eqref{eq:general-formulation} on a subset of tasks $\mathcal{A}_{ID}$ (In Distribution). We then train the static readout weights $\Theta$ of two random RNNs $Y_{\Theta_y}$ and $W_{\Theta_w}$ to replicate, respectively, the functions $Y$ and $W$ on the training data.

\item {\bf Phase 2: The whole architecture is tested (dynamical learning).}
The architecture is tested on a new, previously unseen task $\alpha \in \mathcal{A}_{OOD}=\mathcal{A}\setminus \mathcal{A}_{ID}$ (Out-of-Distribution).
During this phase, all synaptic weights remain fixed. Instead of relying on synaptic plasticity, the virtual weights 
 evolve dynamically according to: $\vb{w}(t+dt) = \vb{w}(t) + W_{\Theta_w}(t)$. 
 The updated output of the network is given by $y(t) = Y_{\Theta_y}(\{\vb{w}(t), \vb{z}(t)\})$. Importantly, we remark that $\vb{w}(t)$ no longer represents physical synaptic weights but rather contextual signals encoded in the activity of neurons, acting as input currents to the $Y_{\Theta_y}$ network.
\end{itemize}

\subsection*{The role of gain-modulation in adaptive behavior}

Recent research has highlighted a growing interest in understanding modulatory mechanisms and their role in neural computation. 
It has been observed that gain modulation allows neurons to adjust their response to inputs based on context \cite{larkum2013cellular}. This dynamic regulation of neuronal sensitivity plays a critical role in adaptive behavior and cognitive flexibility. Moreover, it has been implicated in various functions, including feedback control \cite{meulemans2022least,feulner2022feedback} and attention, where it allows for the selective amplification of relevant information. 
A recent study \cite{jiang2024dynamic} identified gain modulation as a key neural mechanism underlying hierarchical spatiotemporal prediction and sequence learning, demonstrating that such models can replicate various cortical processing features, including temporal abstraction and space-time receptive fields in the primary visual cortex.
Furthermore, gain modulation is closely tied to meta-learning and multitask learning \cite{perez2018film,wybo2023nmda}, enabling neural networks to efficiently generalize across diverse tasks by leveraging contextual cues.

Within biological recurrent networks, dendritic segregation provides a structural basis for gain modulation. By compartmentalizing synaptic inputs, neurons can effectively implement non-linear interactions between different input streams. This is reminiscent of In-Context Learning (ICL) in machine learning, where multiplicative interactions are crucial for enabling models to leverage previously learned information in new contexts.
Although we focus on gain modulation mediated by apical dendrites, other mechanisms are also known. For instance, neuro-glia interactions at the synapse have been proposed as a substrate for working memory, with astrocytes playing a crucial role in modulating neural activity and maintaining memory through glia–synapse interactions \cite{de2022multiple}.
This gain modulation mechanism could complement other adaptive processes, such as subtle, correlated changes in connectivity \cite{feulner2022feedback}. 

Our work explores the role of gain modulation in recurrent networks, employing a simplified model where apical currents act multiplicatively on basal currents within a transfer function. A key parameter, $\gamma$, controls the strength of these non-linear interactions (see Methods for further details). By modulating the gain of neuronal responses, these networks exhibit enhanced adaptability and in-context learning capabilities.

\section*{Results}

\begin{figure*}[ht!]
\centering
\includegraphics[width=\linewidth]{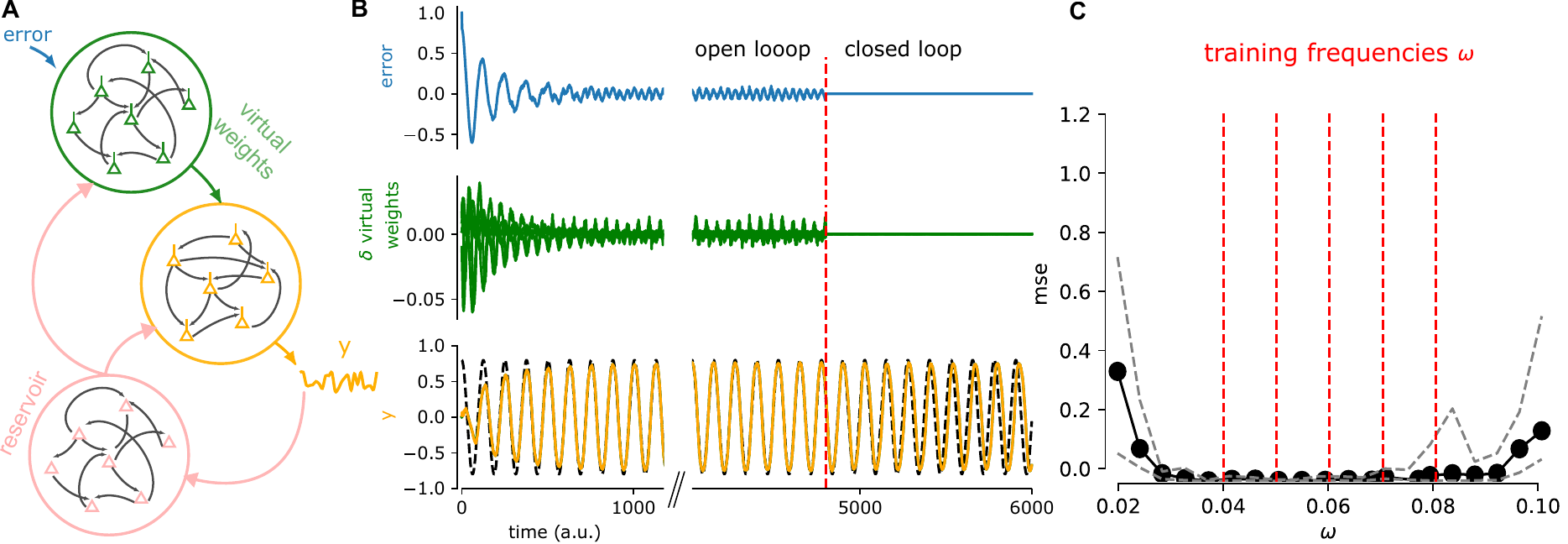}
\caption{ \textbf{Dynamical adaptation of temporal trajectories:}
\textbf{A.} Overview of the network architecture employed. A recurrent network composed of $N=20$ units (described by the vector \textcolor{darkpink}{$\vb{z}(t)$}, depicted in pink, which follows \EqRef{eq:rnn}) is utilized for learning a periodic trajectory, following the prescription of reservoir computing.
One recurrent network is tasked with estimating the gradient of virtual weights, while another is dedicated to estimating the behavioral reconfiguration (illustrated in green and orange, respectively) resulting from these updated virtual weights.
\textbf{B.} Illustration of our architecture dynamically adjusting (without synaptic alterations) to adhere to the desired dynamics. Errors (represented by the blue trajectory) are fed back to the initial network to assess necessary updates to the virtual weights $\delta w$, a $N$-dimensional vector (shown in the green trajectory). Subsequently, these updates are transmitted to the second network, modifying the decoding of reservoir dynamics (indicated by the orange lines). Initially, the reservoir receives the target trajectory as input (in open loop, before the red vertical line), which is later replaced by the estimated trajectory itself (in closed loop, after the red vertical line).
\textbf{C.} Our networks were pre-trained on five target frequencies (marked by red vertical lines) and tested across a range of frequencies, evaluating the mean squared error (MSE) 
between the target and estimated trajectories in closed-loop scenarios (solid: median, dashed: 20-th/80-th percentile (statistics evaluated over 10 realizations).}
\label{fig2}
\end{figure*}

\subsection*{Dynamical adaptation for temporal trajectories}

We consider the task of autonomously predicting a temporal trajectory $\{f(t) = y^{targ}(t)\}_t$. Following the reservoir computing paradigm (see Methods section), we employ a random RNN (Fig. \ref{fig2}A, pink), whose dynamics follows Equation \eqref{eq:rnn}, to extract temporal features $\vb{z}(t)$.
During training, the reservoir receives the target dynamics $y^{targ}(t)$ as input ($x(t) = y^{targ}(t)$ in \EqRef{eq:rnn}) and is tasked with predicting the subsequent step of the trajectory via a linear readout of its activity (open loop, see Fig. \ref{fig2}B). This setup can be reformulated as a dynamical supervised learning problem, as described in the methods section. 
The readout weights can then be dynamically adjusted, minimizing the error between the target and the current prediction $y(t)$.
This results in the following dynamics:

\begin{equation}\begin{cases}
e & =  (y^{targ} - y) \\
y & =   \textcolor{orange}{ Y_{\Theta^y}}(\{\textcolor{darkpink}{\vb{z}},  \vb{ w } \} ) \simeq \textcolor{darkpink}{\vb{z}}\cdot \vb{ w } \\
\tau_w \, \dot{\vb{w}} & =  \textcolor{darkgreen}{ W_{\Theta^w}}(\{\textcolor{darkpink}{\vb{z}}, e\}) \simeq \textcolor{darkpink}{\vb{z}}\, e
\end{cases}
\end{equation}

Here, the network \textcolor{darkgreen}{ $W_{\Theta^w}$}$(\{\textcolor{darkpink}{\vb{z}}, e\})$ is dedicated to estimating the gradient of virtual weights, while \textcolor{orange}{ $Y_{\Theta^y}$}$(\{$\textcolor{darkpink}{$\vb{z}$}$, \vb{w}\})$
is tasked with estimating the predicted $y(t)$ as a function of the new virtual weights (Fig.\ref{fig2}A green and orange respectively). 

These gain-modulated architectures are pre-trained to replicate, respectively, gradient descent updates and scalar products obtained on a set of $N_{train}$ training sequences $\{\{y_\alpha^{targ}(t)\}_t :\ \alpha \in [N_{train}]\}$.
More specifically, the target sequences are $5$ sinusoidal functions with different frequencies (see vertical red lines in \FigRef{fig2}C.) 

In the closed-loop phase, the features \textcolor{darkpink}{$\vb{z}$} are obtained by directly feeding the network estimation $y(t)$ as the input to the RNN ($x = y(t)$ in \EqRef{eq:rnn}). This results in an autonomous dynamical system that reproduces the target trajectory. (see Fig. \ref{fig2}B). 
In Fig.\ref{fig2}B we report an example of successful dynamical adaptation of our model. 
Errors, represented by the blue trajectory, are fed back to the initial network to assess necessary updates to virtual weights, shown in the green trajectory. 
These updates are then transmitted to the second network, modifying the decoding of reservoir dynamics indicated by the orange lines. 
Initially, the reservoir receives the target trajectory as input in an open-loop fashion before the red vertical line, which is later replaced by the estimated trajectory itself in a closed-loop configuration after the red vertical line.
The networks were pre-trained on five target frequencies marked by red vertical lines in Fig.\ref{fig2}C and then tested across a range of frequencies. 
Evaluation was performed by assessing the mean squared error (MSE) between the target and estimated trajectories in closed-loop scenarios (see Fig.\ref{fig2}C, solid and dashed lines represent respectively, median and 20th/80th percentile range. Statistics is evaluated over 10 realizations of the experiment).

\begin{figure}[h!]
\centering
\includegraphics[width=\linewidth]{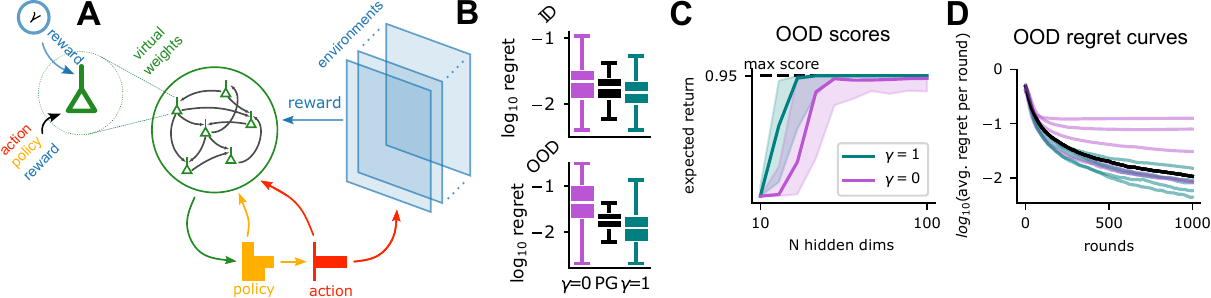}
\caption{ \textbf{Dynamical reinforcement learning: Multi Armed Bandits}. 
\textbf{A.} Schematic of network architecture and task. 
Virtual weighs parameterize the agent's policy (orange). At each round, an action (red) is sampled from the policy and played. 
A reward is then sampled from the current environment. 
A GM-network (green) then predicts the virtual parameter update. 
\textbf{B.} Regret comparison between the distilled and the original policy gradient algorithm. We report log regret per round distribution achieved at the 100-th round for 100 independently trained models in ID and OOD settings. 
We compare a gain-modulated network ($\gamma = 1$), and a network without gain modulation ($\gamma=0$) with the policy gradient training source (PG).
\textbf{C.} OOD model performance varying number of hidden dimensions. Solid lines indicate the median score (expected reward), computed over 100 independently trained models. The filled area indicates 20-80\% confidence interval. 
\textbf{D.} We report average regret per round curves in OOD setting. We compare a gain-modulated network ($\gamma = 1$, teal), and a network without gain modulation ($\gamma=0$, orchid) with the policy gradient training source (in black). Each curve represents the average regret per round for one fixed trained model, computed by averaging over 100 independent simulations.}
\label{fig3}
\end{figure}

\subsection*{Extension to Reinforcement learning and the role of Gain modulation}

Dynamical adaptation is now investigated within the framework of reinforcement learning. We provide robust evidence supporting the hypothesis that gain modulation represents a fundamental component in implementing in-context learning in biological agents.
We first explore stateless environments $\mathcal{A}^{bandits}$, which are represented by Bernoulli K-armed bandits \cite{berry1985bandit}. 
Within each environment $\alpha\in \mathcal{A}^{bandits}$,
there exists a subset $P_\alpha\subset [K]$ of arms that yield a reward with high probability ($p=0.95$). 
During the training phase, the in-distribution environments give a high reward probability to even-numbered arms, whereas during testing, the out-of-distribution environments assign a high reward probability to odd-numbered arms.

In this simplified bandit scenario, where state information is absent, the virtual weights $\vb{w}$ directly parameterize the policy probabilities: $\vb{\pi} = \mathrm{softmax}{(\vb{w})}$. 
Consequently, our focus lies primarily on analyzing the behavior of the network \textcolor{darkgreen}{$W_{\Theta^w}^{\gamma}$}$( \cdot )$ acting on the virtual parameters (\FigRef{fig3}A, green). 
This serves as an ideal test bed for evaluating the network's ability to learn the policy gradient update rule and generalize it to out-of-distribution scenarios.
The network \textcolor{darkgreen}{$W_{\Theta^w}^{\gamma}$}$( \vb{\pi}, a, r )$ has a gain-modulated architecture, as described in the Methods section, and is a function of the reward $r$ and the action $a$. Here $\gamma$ indicates the dependence on the gain-modulation strength.
We train the network to approximate the policy gradient update rule using policy gradient estimates from a single learning trajectory (1000 rounds, learning rate $lr=0.1$) in an in-distribution environment. We then test the distilled policy gradient networks with a higher learning rate ($lr=1.0$) in out-of-distribution (OOD) environments.
To systematically investigate the impact of gain modulation ($\gamma = 1$) on out-of-distribution performance, we compare it with networks where the reward does not modulate the network gain ($\gamma = 0$). For each case, we select the optimal hyperparameters through a grid-based search (additional details in the Appendix).

Our experimental findings are presented in Fig.\ref{fig3}. First, comparing models of different sizes (number $N$ of hidden units), we find that models with gain modulation require significantly fewer neurons to achieve maximum scores in OOD environments (Fig.\ref{fig3}C).
In Fig.\ref{fig3}B, we report the regret per round distribution at the 100th round and compare it with the distribution obtained by policy gradient with the same learning rate and iterations. A gain-modulated network achieves a regret distribution comparable to the policy gradient in both ID and OOD environments, with a lower median. In contrast, networks without gain modulation show significantly higher regret.
Analyzing the regret curves in Fig.\ref{fig3}D, we observe that a model with gain modulation often learns a more data-efficient algorithm than its source, even in OOD environments. Conversely, a model without gain modulation fails to generalize to OOD environments and does not converge to zero regret.

In summary, gain modulation enables the network to consistently and efficiently distill the correct gradient update rule and generalize it to unseen environments, predicting the correct virtual weight update in regions of the input space far from the training data.

\begin{figure*}[t!]
\centering
\includegraphics[width=\linewidth]{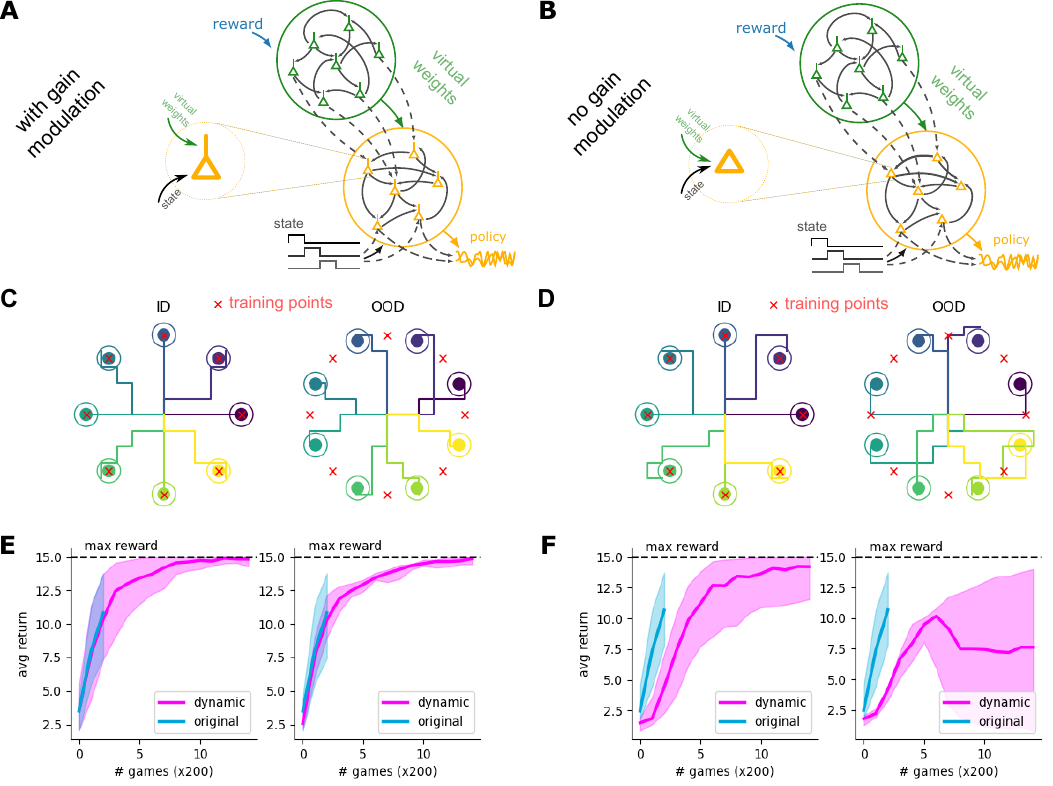}
\caption{\textbf{Dynamical Reinforcement Learning Dark Room}
\textbf{A.} Overview of the network architecture employed. One GM-network (depicted in green) is responsible for estimating the necessary update of virtual weights through policy gradient, while another GM-reservoir is focused on estimating policy reconfiguration (shown in orange) resulting from changes in virtual weights.
\textbf{C.} Illustration of the task: the agent begins from the center and learns to reach not observable objects positioned at various locations (represented by colored circles). After several trials, the agent achieves precise targeting of the circles.
\textbf{C.} Reward as a function of number of trials (or games), blue: policy gradient, left panel pink: dynamical learning for training positions (ID), right panel pink: dynamical learning for new positions (OOD).
Line: median, shading: 20-th/80-th percentile range.
\textbf{B, D, F.} Similar to \textbf{A, C, E}, respectively, but without gain modulation.
}
\label{fig4}
\end{figure*}

\subsection*{Dynamical Reinforcement learning in a reaching task}

We examine a reaching task, known as the dark room task (a simple instance of the water maze task \cite{morris1981spatial}), set in a 2D maze within the domain $(-1,1)\times(-1,1)$ with a grid size of $0.1$. 
Within this grid-like environment, the agent navigates by selecting one of four actions: up, down, left, or right, thereby determining its subsequent position. 
The primary goal is to locate a concealed object within the maze, with the agent having sole awareness of its own position. 
Feedback is provided via rewards, where the agent receives a reward of $10$ if the distance to the object is $0.1$, $15$ if the distance is $0$, and $0$ otherwise. 
Through iterative exploration, the agent develops a strategy to efficiently traverse the maze and pinpoint the object despite the limited visibility.
The position of the agent is encoded separately using 25 input units $\vb{x}$ each, employing Gaussian activation functions distributed on a 5x5 grid in the maze and with a width of $0.2$.

To accomplish this task we consider a network architecture composed of two networks. 
One GM-network, \textcolor{darkgreen}{$W_{\Theta^w}^{\gamma}$}$(\{\vb{z}, r, a,\vb{\pi}\}) $ is responsible for estimating the necessary update of virtual weights through policy gradient (\FigRef{fig4}A, green network), while another GM-network \textcolor{orange}{$Y_{\Theta^y}^{\Gamma}$}$(\{\vb{ z },  \vb{ w }\}) $ estimates policy reconfiguration (\FigRef{fig4}A, shown in orange) resulting from changes in virtual weights. Here, the parameters $\gamma, \Gamma$ define the strength of the gain modulation, as explained in the Methods section.
Their dynamics can be described by the following equation Eq.\eqref{eq:dyn-pg}.
Indeed, we compare the case with  (\FigRef{fig4}A) and without (\FigRef{fig4}B) gain modulation.
We refer to Supporting Information for further details on the training procedure.

Firstly, we assessed the performance of the policy gradient algorithm using a set of 8 food locations (refer to Fig. \ref{fig4}C, indicated by red crosses), where the total reward averaged over 200 trials was observed against the number of trials (\FigRef{fig4}E, depicted by blue lines, thin lines for individual positions and thick lines for the average across all positions). After multiple trials, the agent successfully achieved precise targeting of the circles. 
Data collected from these experiments were utilized to train our networks to estimate \textcolor{darkgreen}{gradients} and \textcolor{orange}{scalar products}, as defined in \EqRef{eq:dyn-pg}.

To validate that our trained model is capable of dynamically implementing policy gradient itself, we tested it on both the training set locations (ID, \FigRef{fig4}C, left panel) and new test positions (OOD, \FigRef{fig4}C, right panel). 
For each food position (coded with different colors), we illustrate a sample trajectory executed by the agent (in corresponding colors) to reach the target at the end of the training. 
The agent's precision closely matches that of the plastic policy gradient learning rule. 
We present the reward plotted against the number of trials for both training (\FigRef{fig4}E, right panel, pink line) and test (\FigRef{fig4}C, right panel, pink line) food locations.

We compared these performances against an architecture lacking gain modulation ($\gamma=0$), observing worst performances (see \FigRef{fig4}B, D, F). 
Notably, while performances for ID food locations are acceptable (\FigRef{fig4}E, left panel, pink lines), those for OOD cases are extremely poor (\FigRef{fig4}E, right panel, pink lines).
This observation, coupled with the results from the preceding section, suggests that gain modulation is a crucial component in facilitating the generalization of adaptive behavioral capabilities in recurrent networks.

\paragraph{Temporal credit assignment and Future Reward}

We demonstrate, that our architecture is capable to perform the temporal computation required to learn delayed action-reward temporal relations (see section \ref{discount_factor_appendix} in Appendix), requiring evaluating temporal credit assignment. 
Indeed, in reinforcement Learning (RL) an agent learns to maximize cumulative rewards. The discount factor, usually denoted by $\gamma_d$ (where $0 \leq \gamma_d \leq 1$), is crucial in RL as it determines the importance of future rewards. A higher $\gamma_d$ values future rewards more significantly, encouraging the agent to consider the long-term consequences of its actions. Conversely, a lower $\gamma_d$ makes the agent prioritize immediate rewards. The cumulative reward $R_t$ at time step $t$ is given by:
\begin{equation}
R_t = \sum_{s \geq t} r_s \gamma^{(s-t)} _d
\end{equation}
To account for this, the term $\vb{e}$ in Eq.\eqref{eq:dyn-pg} should be replaced by its exponential temporal filtering, with a timescale $\tau_e=-\frac{1}{\log(\gamma_d)}$ (see also Eq.\eqref{rl_discount} in the Appendix). In this way actions in the past, can be potentiated when delayed rewards are received.

If the discount factor is zero, the agent's policy would change only when it is very close to the moment the reward is received. In a reaching task, the agent will only change its policy when it is very close to the reward, completely ignoring the need for long-term planning and making it unlikely to ever reach the goal from distant starting points (see \FigRef{figs2}.D-E). As a result, the policy points towards the target position only when nearby the target itself (see \FigRef{figs2}F), resulting in failure when the agent randomly moves in the wrong direction at the beginning of the task (see \FigRef{figs2}F, black line).

On the other hand, recurrent connections enable a network to maintain a memory of past states and actions, effectively allowing it to use information from previous time steps to inform current decisions, achieving optimal planning and decisions even when far from the target location (and hence, from the reward, see \FigRef{figs2}.A-B-C). 

\subsubsection*{Application to a realistic task}

We test our framework to a more complex and realistic setting by applying it to a MuJoCo ant environment, demonstrating its scalability to continuous control tasks. The agent, modeled as a quadruped, operates in a high-dimensional control space with 27 environmental state variables and 8 control torques (2 per leg), requiring coordinated control of multiple degrees of freedom.

\begin{figure*}[t!]
\centering
\includegraphics[width=0.9\linewidth]{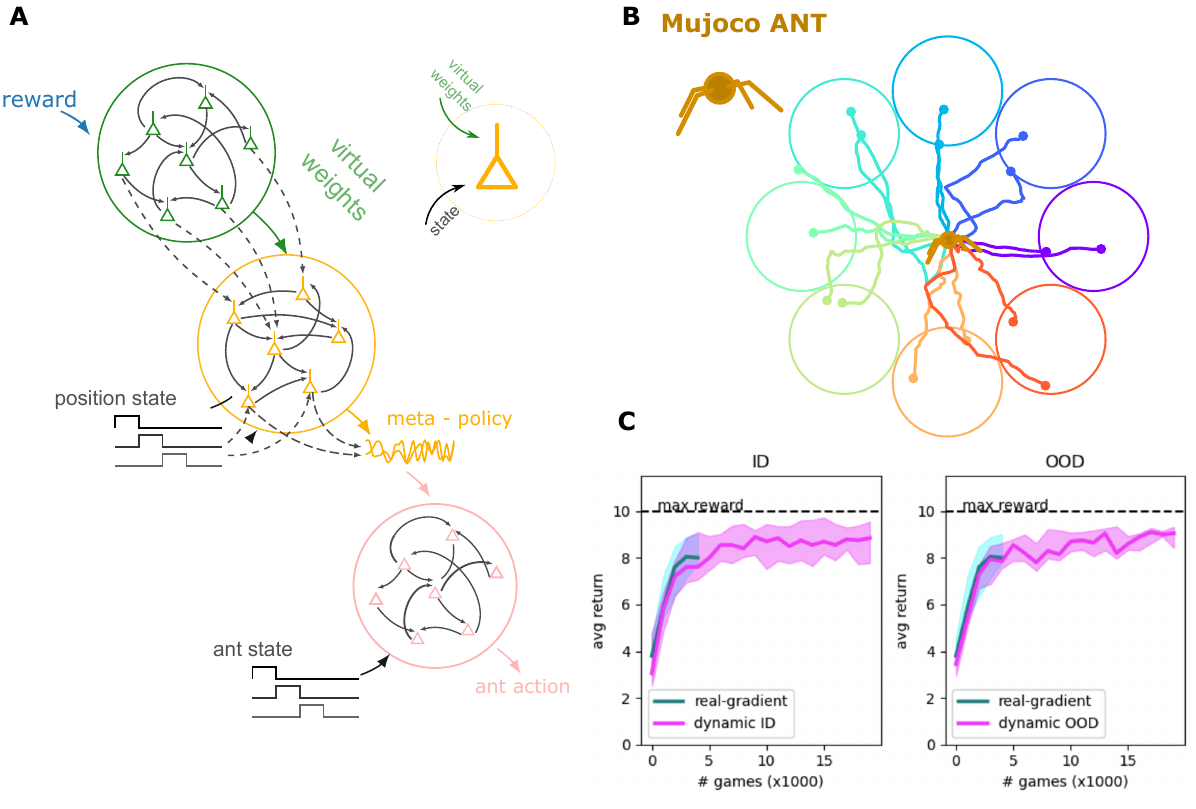}
\caption{\textbf{Dynamical Reinforcement Learning Dark Room}
\textbf{A.} 
Overview of the network architecture employed. One GM-network (depicted in green) is responsible for estimating the necessary update of virtual weights through policy gradient, while another GM-reservoir is focused on estimating policy reconfiguration (shown in orange) resulting from changes in virtual weights. 
\textbf{B.} Illustration of the task, and of the mojoco environment: the agent begins from the center and learns to reach not observable objects positioned at various locations (represented by colored circles). After several trials, the agent achieves precise targeting of the circles. Our networks are pre-trained on a set of target points (red crosses) an then tested on the same (displayed in the left panel) and new positions (shown in the right panel).
\textbf{C.} 
Reward as a function of number of trials (or games), blue: policy gradient, left panel pink: dynamical learning for training positions (ID), right panel pink: dynamical learning for new positions (OOD).
Line: median, shading: 20-th/80-th percentile range.
}
\label{fig5}
\end{figure*}

To manage low-level motor execution, we first trained a network to perform four fundamental motor primitives—moving left, right, forward, and backward. This pretraining step allows higher-level adaptation without direct joint-level optimization (see Appendix for details). Building on this, we use our gain-modulated network as a meta-policy that dynamically selects motor primitives based on the agent’s position. The meta-policy network (orange network \FigRef{fig5}A) receives positional inputs and determines the appropriate primitive, while a secondary network (green network \FigRef{fig5}A) modulates its behavior through virtual weights, mirroring the architecture used in the grid-based task.

This architecture enables flexible and efficient adaptation: the agent successfully generalizes to target locations not included in the training set (see \FigRef{fig5}C), adjusting its movement strategies dynamically without modifying its underlying synaptic structure. This capability is particularly significant for neuromorphic computing and robotics, where adaptive behavior is often constrained by the need for on-chip plasticity. Our results demonstrate that a fixed neuromorphic chip, programmed with a gain-modulated network, can dynamically adjust to new goals without requiring structural modifications. This approach reduces computational overhead and power consumption while enabling efficient, low-power autonomous systems that adapt in real time. By eliminating the need for weight updates at the hardware level, this architecture provides a scalable and robust foundation for adaptive physical agents, paving the way for next-generation neuromorphic robotics.

\section*{Discussion}

The brain's ability to flexibly and rapidly adapt to novel tasks remains only partially understood. Different studies have highlighted that this adaptability relies on mechanisms such as dynamic network reconfiguration and context-dependent modulation, allowing agents to respond flexibly to environmental changes. However, a significant gap remains in understanding how these processes map onto specific neural substrates.

While synaptic plasticity has been extensively studied as a mechanism underlying learning and memory \cite{abraham2019plasticity}, emerging evidence suggests that it alone may not account for the full spectrum of adaptive behavior. Abraham et al. critically evaluate the synaptic plasticity and memory (SPM) hypothesis, concluding that non-synaptic mechanisms, such as intrinsic plasticity, structural neural changes, epigenetic regulation, glial contributions, network reorganization, and protein turnover, likely contribute significantly to memory storage and retrieval, thus broadening the framework for understanding dynamic adaptation. This perspective questions long-standing assumptions about neural circuits, highlighting that we may still lack full understanding of many processes we consider well-established.

Inspired by the brain's adaptability, artificial neural network architectures have been challenged to replicate similar capabilities. Fast synaptic plasticity, hypothesized to orchestrate flexible cognitive functions, inspired several artificial architectures, such as fast-weight networks \cite{schmidhuber1992learning, irie2021going} and hyper-networks that predict main network weights \cite{ba2016using}. However, these approaches often rely on biologically unrealistic mechanisms, such as requiring neurons to directly act on the synapses of a second network or implementing "inner loops" in network dynamics \cite{chalvidal2022meta, ba2016using}. More biologically plausible alternatives, like coupled neural-synaptic dynamics \cite{miconi2018differentiable}, come closer to natural systems but still depend on complex credit assignment mechanisms.

Transformers, known for their remarkable in-context learning (ICL) capabilities, excel at tasks requiring adaptability, such as natural language and image processing \cite{vaswani2017attention, achiam2023gpt, ramesh2021zero}. Mechanisms such as attention, associative memory \cite{ramsauer2020hopfield}, and induction-based copying by attention heads \cite{olsson2022context} have been proposed to explain ICL in transformers, with theoretical work showing that attention mechanisms can implement gradient updates \cite{von2023transformers}. However, the emergence of these capabilities remains poorly understood, particularly when comparing them to the in-context learning observed in biological systems.

We propose a constructive method to induce in-context learning in biologically plausible neural networks, in a broad variety of scenarios.
We offer an alternative formulation in which dynamic adaptation is modulated by dendritic input segregation and amplification. 
Virtual (fast) weights are encoded by the activity of a second network that modulates the transfer function of neurons in the base network. 
As a result, virtual weights are not synaptic weights, but their values and updates are evaluated and encoded by the activity of other neurons, which were pre-trained to perform gradient updates. Another pre-trained network receives those weights signals as an input, along with the current environment features, and computes an adapted in-context response. 

This mechanism overcomes the strict locality constraints of synaptic plasticity, which require all elements involved in evaluating weight updates to be present at the synaptic site. Consequently, biologically plausible synaptic updates cannot explicitly rely on information from other weights, a requirement often seen in backpropagation methods. In contrast, virtual weights are encoded by neuronal activity and can propagate across the neural circuit, bypassing these limitations.
This allows virtual weight updates to depend on other virtual weights, enabling a more natural and precise evaluation of credit assignment. 

Our work generalizes and provides a biologically plausible implementation for deep gated RNNs performing ICL \cite{zucchet2023gated}, maintaining a similar number of trainable parameters. By dividing the architecture into two components and introducing an additional objective that forces one network to approximate the gradient of the other, we achieve robust dynamic adaptation in tasks requiring temporal learning. 
When tasked to learn temporal trajectories, through error feedback and virtual weight updates, our network achieves successful dynamical adaptation, without synaptic plasticity. Similarly, in reinforcement learning scenarios, the policy in response to the environment state is dynamically modulated by virtual weights, which are updated in function of the reward.
We find that networks with gain modulation exhibit improved performance and robustness compared to those without gain modulation. Moreover, these gain-modulated networks demonstrate more data-efficient learning algorithms, outperforming counterparts in both in-distribution and out-of-distribution environments, showing their capability to face novel scenarios.

Our findings demonstrate that adaptive network architectures can emerge by selecting appropriate readout weights in initially random networks, without requiring explicit backpropagation. These weights can be shaped through biologically plausible mechanisms such as evolutionary processes, slow reinforcement learning, or local plasticity rules relying on feedback alignment. Unlike standard deep-learning models—such as transformers, gated RNNs, and hyper-networks—that depend on non-local credit assignment mechanisms like backpropagation through time, our approach leverages local, dynamically reconfigurable processes, making it more compatible with biological learning principles.

In line with this, our model suggests that dynamic reconfiguration provides a biologically plausible mechanism for rapid adaptation, complementing traditional fast synaptic plasticity. Different neural circuits in the brain may rely on varying balances of dynamic reconfiguration and synaptic plasticity, shaped by their functional roles and evolutionary pressures. In the Supplementary Materials, we provide a proof of concept through a T-maze simulated experiment, where we observe a substantial difference in how contextual information is encoded for synaptic versus adaptive learning. This distinction supports the potential for experimental validation.

In conclusion, our approach provides an explicit framework to induce in-context learning (ICL) in biologically plausible networks, offering a constructive pathway toward understanding ICL in biological systems and bridging the gap between neuroscience and machine learning.

\section*{Methods}

\subsection*{Gain modulated reservoir computing (GM-RC)}
\paragraph{Reservoir Computing} 
Reservoir Computing (RC) represents a paradigm in machine learning that provides a model for biologically plausible computation and learning, drawing inspiration from the information processing mechanisms observed in biological neural networks.

In this approach, a random RNN is employed to extract features from a time-dependent signal $\vb{x}(t)$. The dynamics of the RNN describe the evolution of $N$ hidden units $\vb{z}(t) = (z_1(t),\cdots,z_N(t))$, governed by the following differential equation: 
\begin{align}\label{eq:rnn}
    \tau_z\dot{\vb{z}} = \phi \left( J \vb{z} + R\vb{x} \right) - \vb{z}
\end{align}
The value of each unit represents the activity of a population of neurons following the Wilson and Cowan formulation (\cite{wilson1972excitatory}).
Here, $J$ and $R$ represent fixed random matrices, representing the recurrent connections and the projection from inputs to hidden units, respectively.
Subsequently, the features extracted by the network can be utilized to predict a target signal $\vb{y}^{targ}(t)$ by learning readout weights $\Theta$, such that $|\vb{y}^{targ}(t) - \Theta\vb{z}(t)|^2$ is minimized.
The reservoir is then implementing a map 
\begin{align}
    \vb{y}(t) = \mathrm{RNN}_\Theta(\{\vb{x}(s)\}_{s\le t}) 
\end{align}
where $\Theta$ represents trainable parameters. In the rest of the article, we will write  $\vb{y} = \mathrm{RNN}_\Theta(\{\vb{x}\})$, dropping the temporal dependencies.  
\footnote{
For this mapping to be well-defined, we can assume the RNN to have the echo-state property \cite{yildiz2012re}.
}

When addressing input-output mappings without time dependencies, we consider a network operating in the $\tau_z, J_{ij} \to 0$ limit. In this scenario, the network computes an instantaneous function of the input, represented as $\mathrm{NN}_\Theta(x) = \Theta\phi\left( R\vb{x}\right)$. Essentially, this is equivalent to considering a one-layer feed-forward network with random fixed input weights. 
This architecture is also referred to in the literature as the Extreme Learning Machine \cite{huang2006extreme}.

\paragraph{Gain modulated network architecture} Analysing Equations \eqref{eq:dyn-gd} and \eqref{eq:dyn-pg} we observe that a network must possess the capability to perform multiplications of its inputs to approximate a gradient descent update of virtual parameters.
Building upon this insight, we introduce a gain-modulated reservoir network (GM-RC). 
This architecture draws inspiration from the morphology and function of pyramidal neurons in the cortex, which nonlinearly integrate inputs from basal and apical dendrites \cite{larkum2013, shai2015physiology}. 
Here, we consider an additional input source $\vb{x}^{\text{ap}}$ randomly projected into the apical dendrite of each neuron by the matrix $R^{\text{ap}}$. 
Consistent with experimental observations in L5 pyramidal neurons, we allow the apical inputs to modulate the gain of the activation function, thereby altering its slope.
Consequently, the resulting RNN equation is formulated as:
\begin{align}
\tau_z \dot{\vb{z}} &= \phi \left( \left(\vb{b}^{\text{ap}}+ \gamma \cdot R^{\text{ap}}\vb{x}^{\text{ap}}\right)\odot (J \vb{z} + \beta\cdot R^{\text{ap}}\vb{x}^{\text{ap}} + R\vb{x}) \right) - \vb{z}\
\end{align}
Where $\vb{b}^{\text{ap}}$ is a constant bias vector. The hyperparameters $\beta, \gamma$ modulate the effect of the gain modulation of the apical inputs.  Specifically, when $\gamma = 0$, we obtain a network in which $\vb{x}^{\text{ap}}$ does not affect the gain modulation. 
Similarly, as before, the expression $\vb{y} = \mathrm{GMRNN}^\gamma_\Theta(\{\vb{x}| \vb{x}^{\text{ap}}\})$ will denote the input $\to$ output mapping implemented by a gain-modulated reservoir. As before, removing time dependencies, we will have an instantaneous function $\vb{y} = \mathrm{GMNN}^\gamma_{\Theta}(\vb{x}|\vb{x}^{\text{ap}}) = \Theta \phi \left( \left(\vb{b}^{\text{ap}} + \gamma \cdot R^{\text{ap}}\vb{x}^{\text{ap}}\right)\odot (R\vb{x}) \right)$ of the inputs $\vb{x}^{\text{ap}}$ and $\vb{x}$.
We explicitly maintain the dependence on $\gamma$ because, in our experiments, we use this parameter to regulate the gain modulation effect of the apical inputs on the network.
The choice for the name $\vb{x}^{\text{ap}}$ is inspired by the current received in the apical dendrites of L5 pyramidal neurons, that are believed to carry contextual/high-level information \cite{larkum2013}.

\subsection*{Dynamical supervised learning for temporal trajectory}

We consider the task of learning a target temporal trajectory. Each environment $\alpha$ is defined by a target trajectory $y_\alpha^{targ}(t)$, such that the input and the feedback $x_\alpha(t) = f_\alpha(t) = y_\alpha^{targ}(t)$ are both given by the target sequence. The network response is defined as the current estimation of $y^{targ}(t)$ via a linear readout $y(t) = \vb{ w } \cdot \vb{z}(t)$, where $\vb{z}$ are the activity of a random RNN that operates as a reservoir computer extracting nonlinear features of the input sequence. The RNN dynamics is given by equation \eqref{eq:rnn}, with input $x$ given by the target sequence. The readout weights in turn dynamically adapt by following the delta rule, minimizing the reconstruction error. 

The learning process can then be described by the following dynamical system:
\begin{equation}\label{eq:dyn-deltarule}
\begin{cases}
e & =  (y^{targ} - y) \\
y & =  \vb{ z } \cdot \vb{ w } \\
\tau_z\dot{\vb{z}} & = \phi \left( J \vb{z} + R\vb{x} \right) - \vb{z}\\
\tau_w \, \dot{\vb{w}} & =  \vb{z}\, e
\end{cases}
\end{equation}
where we removed the dependence on time $t$ for simplicity.
The operations required are nonlinear, and usually are viewed as implemented by the plasticity rule and by the multiplication between presynaptic activity and synaptic weights. However, here we interpret $\vb{w}$ as dynamic variables, proposing, as a possible implementation, that such non-linear functions are computed by two neural networks  \textcolor{darkgreen}{$W_{\Theta^w}$}, and \textcolor{orange}{$Y_{\Theta^y}$}:
\begin{equation}\begin{cases}\label{eq:dyn-gd}
e & =  (y^{targ} - y) \\
\tau_z\dot{\vb{z}} & = \phi \left( J \vb{z} + R\vb{x} \right) - \vb{z}\\
y & =   \textcolor{orange}{Y_{\Theta^y}}(\{\vb{x}| \vb{w}\}) \simeq Y(\vb{x},\vb{ w }) = \vb{x} \cdot \vb{w} \\
\tau_w \, \dot{\vb{w}} & =  \textcolor{darkgreen}{W_{\Theta^w}}(\{\vb{x}| e\}) \simeq W(\vb{x}, e) = \vb{x}\, e
\end{cases}
\end{equation}
In particular, we used two GMRNNs that receive $\vb{x}, \vb{ w }, e$ as inputs and provide the proper output thanks to a suited training of their readout weights ($\Theta^y$ and $\Theta^w$, following reservoir computing paradigm).
Notice that now \eqref{eq:dyn-gd} represents an RNN dynamically performing the learning process.

\subsection*{Dynamical reinforcement learning}
In the context of reinforcement learning, our framework outlines a systematic approach for modelling agents that can dynamically adapt their behavior across diverse environments. 
We assume the same state transition dynamics across all the environments $\dot{\vb{x}} = X(\vb{x}, a)$, while the reward distribution 
$r \sim Prob_\alpha(r|\vb{x}, a)$ depends on the current environment $\alpha \in \mathcal{A}$. Here $a(t)$ is an integer value indicating the action at time $t$ among $\mathrm{D}$ possible ones.

We start by defining a policy network, denoted as $\vb{\pi} =\mathrm{softmax}(\vb{y})$
which implements a policy mapping the agent state encoded by the feature vector $\vb{z}$, to a probability distribution over actions. 
We assume an agent that linearly recombines these features $\vb{y} = Y(\vb{z}, \vb{w}) = \vb{w}\cdot \vb{z}$. 
This assumption is made without loss of generality, since we could always employ a reservoir computer as an intermediate feature extraction layer, as described in the previous subsection.
The agent learns by evaluating the policy gradient with respect to the weights $\vb{w}$. 
This leads to a dynamics similar to equation \eqref{eq:dyn-gd},  where the definition of $e(t)$ is changed as follows:
\begin{equation}
\label{eq:pg-nodisc}
\begin{cases}
\vb{e} & =   r \left(\vb{\mathds{1}}_a - \vb{\pi} \right) \\
\vb{y} & =  \vb{w} \cdot \vb{z} \\
\tau_w \, \dot{\vb{w}} & =  \vb{e} \odot \vb{z}\, 
\end{cases}
\end{equation}
Here $\mathds{1}_{a(t)}$ represents the 'one-hot encoded' action (as defined in \cite{bellec2020,capone2022towards}) at time $t$. It is a $\mathrm{D}$-element vector where the $a(t)$-th element is one, and all other elements are zero.
This formulation holds for a null discount factor \cite{bellec2020,capone2022towards}), we refer to the Appendix section \ref{discount_factor_appendix} for the description of the general case.

As before, we model that the policy is modulated by an auxiliary population of virtual weights $\vb{w}$, determined by the activity of an additional GMRNN $W_{\Theta_y}$.
This auxiliary network adjusts its internal activity based on the rewards received, effectively implementing policy gradient updates of the virtual weights and thereby modulating the agent behavior in real-time.
Then, the above dynamics can be approximated by two networks, one for estimating the \textcolor{darkgreen}{gradients}, and one for the \textcolor{orange}{scalar product}.
\begin{equation}\label{eq:dyn-pg}\begin{cases}
\tau_w \, \dot{\vb{w}} & = \textcolor{darkgreen}{W_{\Theta^w}}(\{\vb{z} |r,a,\vb{\pi}\}) \simeq  W(\vb{z},r,a,\vb{\pi}) =  r \left(\vb{\mathds{1}}_a - \vb{\pi}(\vb{y}) \right) \odot \vb{z}\, \\
\vb{ y } & = \textcolor{orange}{Y_{\Theta^y}}(\{\vb{ z }|  \vb{ w }\}) \simeq Y(\vb{ z },  \vb{ w }) = \vb{ x }  \cdot \vb{ w }
\end{cases}
\end{equation}


\section*{Acknowledgments}
This research has received financial support from the Italian National Recovery and Resilience Plan (PNRR), M4C2, funded by the European Union - NextGenerationEU (Project IR0000011, CUP B51E22000150006, `EBRAINS-Italy').
LF aknowledges support by ICSC – Centro Nazionale di Ricerca in High Performance Computing, Big Data and Quantum Computing and Sapienza University of Rome (AR12419078A2D6F9).
Work partially funded by the European Union – NextGenerationEU and the Ministry of University and Research (MUR), Italian National Recovery and Resilience Plan (PNRR), M4C2I1.3, project ‘MNESYS’ (PE00000006).
We sincerely thank Maurizio Mattia for helpful discussions on the topic.

\section*{Source code availability}
The source code is available under CC-BY license in the \href{https://github.com/cristianocapone/ABSS}{https://github.com/cristianocapone/ABSS} public repository.

\bibliography{references.bib}

\begin{thebibliography}{10}
\urlstyle{rm}
\expandafter\ifx\csname url\endcsname\relax
  \def\url#1{\texttt{#1}}\fi
\expandafter\ifx\csname urlprefix\endcsname\relax\def\urlprefix{URL }\fi
\expandafter\ifx\csname doiprefix\endcsname\relax\def\doiprefix{DOI: }\fi
\providecommand{\bibinfo}[2]{#2}
\providecommand{\eprint}[2][]{\url{#2}}

\bibitem{lee2013contextual}
\bibinfo{author}{Lee, I.} \& \bibinfo{author}{Lee, C.-H.}
\newblock \bibinfo{journal}{\bibinfo{title}{Contextual behavior and neural
  circuits}}.
\newblock {\emph{\JournalTitle{Frontiers in neural circuits}}}
  \textbf{\bibinfo{volume}{7}}, \bibinfo{pages}{84} (\bibinfo{year}{2013}).

\bibitem{daw2005uncertainty}
\bibinfo{author}{Daw, N.~D.}, \bibinfo{author}{Niv, Y.} \&
  \bibinfo{author}{Dayan, P.}
\newblock \bibinfo{journal}{\bibinfo{title}{Uncertainty-based competition
  between prefrontal and dorsolateral striatal systems for behavioral
  control}}.
\newblock {\emph{\JournalTitle{Nature neuroscience}}}
  \textbf{\bibinfo{volume}{8}}, \bibinfo{pages}{1704--1711}
  (\bibinfo{year}{2005}).

\bibitem{mcdougle2015explicit}
\bibinfo{author}{McDougle, S.~D.}, \bibinfo{author}{Bond, K.~M.} \&
  \bibinfo{author}{Taylor, J.~A.}
\newblock \bibinfo{journal}{\bibinfo{title}{Explicit and implicit processes
  constitute the fast and slow processes of sensorimotor learning}}.
\newblock {\emph{\JournalTitle{Journal of Neuroscience}}}
  \textbf{\bibinfo{volume}{35}}, \bibinfo{pages}{9568--9579}
  (\bibinfo{year}{2015}).

\bibitem{o2017learning}
\bibinfo{author}{O'Doherty, J.~P.}, \bibinfo{author}{Cockburn, J.} \&
  \bibinfo{author}{Pauli, W.~M.}
\newblock \bibinfo{journal}{\bibinfo{title}{Learning, reward, and decision
  making}}.
\newblock {\emph{\JournalTitle{Annual review of psychology}}}
  \textbf{\bibinfo{volume}{68}}, \bibinfo{pages}{73--100}
  (\bibinfo{year}{2017}).

\bibitem{braun2015dynamic}
\bibinfo{author}{Braun, U.} \emph{et~al.}
\newblock \bibinfo{journal}{\bibinfo{title}{Dynamic reconfiguration of frontal
  brain networks during executive cognition in humans}}.
\newblock {\emph{\JournalTitle{Proceedings of the National Academy of
  Sciences}}} \textbf{\bibinfo{volume}{112}}, \bibinfo{pages}{11678--11683}
  (\bibinfo{year}{2015}).

\bibitem{standage2023whole}
\bibinfo{author}{Standage, D.~I.} \emph{et~al.}
\newblock \bibinfo{journal}{\bibinfo{title}{Whole-brain dynamics of human
  sensorimotor adaptation}}.
\newblock {\emph{\JournalTitle{Cerebral Cortex}}}
  \textbf{\bibinfo{volume}{33}}, \bibinfo{pages}{4761--4778}
  (\bibinfo{year}{2023}).

\bibitem{annurev:/content/journals/10.1146/annurev-neuro-092322-100402}
\bibinfo{author}{Heald, J.~B.}, \bibinfo{author}{Wolpert, D.~M.} \&
  \bibinfo{author}{Lengyel, M.}
\newblock \bibinfo{journal}{\bibinfo{title}{The computational and neural bases
  of context-dependent learning}}.
\newblock {\emph{\JournalTitle{Annual Review of Neuroscience}}}
  \textbf{\bibinfo{volume}{46}}, \bibinfo{pages}{233--258},
  \doiprefix\url{https://doi.org/10.1146/annurev-neuro-092322-100402}
  (\bibinfo{year}{2023}).

\bibitem{heald2021contextual}
\bibinfo{author}{Heald, J.~B.}, \bibinfo{author}{Lengyel, M.} \&
  \bibinfo{author}{Wolpert, D.~M.}
\newblock \bibinfo{journal}{\bibinfo{title}{Contextual inference underlies the
  learning of sensorimotor repertoires}}.
\newblock {\emph{\JournalTitle{Nature}}} \textbf{\bibinfo{volume}{600}},
  \bibinfo{pages}{489--493} (\bibinfo{year}{2021}).

\bibitem{heald2023contextual}
\bibinfo{author}{Heald, J.~B.}, \bibinfo{author}{Lengyel, M.} \&
  \bibinfo{author}{Wolpert, D.~M.}
\newblock \bibinfo{journal}{\bibinfo{title}{Contextual inference in learning
  and memory}}.
\newblock {\emph{\JournalTitle{Trends in cognitive sciences}}}
  \textbf{\bibinfo{volume}{27}}, \bibinfo{pages}{43--64}
  (\bibinfo{year}{2023}).

\bibitem{vinyals2016matching}
\bibinfo{author}{Vinyals, O.}, \bibinfo{author}{Blundell, C.},
  \bibinfo{author}{Lillicrap, T.}, \bibinfo{author}{Wierstra, D.} \emph{et~al.}
\newblock \bibinfo{journal}{\bibinfo{title}{Matching networks for one shot
  learning}}.
\newblock {\emph{\JournalTitle{Advances in neural information processing
  systems}}} \textbf{\bibinfo{volume}{29}} (\bibinfo{year}{2016}).

\bibitem{romera2015embarrassingly}
\bibinfo{author}{Romera-Paredes, B.} \& \bibinfo{author}{Torr, P.}
\newblock \bibinfo{title}{An embarrassingly simple approach to zero-shot
  learning}.
\newblock In \emph{\bibinfo{booktitle}{International conference on machine
  learning}}, \bibinfo{pages}{2152--2161} (\bibinfo{organization}{PMLR},
  \bibinfo{year}{2015}).

\bibitem{brown2020language}
\bibinfo{author}{Brown, T.} \emph{et~al.}
\newblock \bibinfo{journal}{\bibinfo{title}{Language models are few-shot
  learners}}.
\newblock {\emph{\JournalTitle{Advances in neural information processing
  systems}}} \textbf{\bibinfo{volume}{33}}, \bibinfo{pages}{1877--1901}
  (\bibinfo{year}{2020}).

\bibitem{singh2024transient}
\bibinfo{author}{Singh, A.} \emph{et~al.}
\newblock \bibinfo{journal}{\bibinfo{title}{The transient nature of emergent
  in-context learning in transformers}}.
\newblock {\emph{\JournalTitle{Advances in Neural Information Processing
  Systems}}} \textbf{\bibinfo{volume}{36}} (\bibinfo{year}{2024}).

\bibitem{vaswani2017attention}
\bibinfo{author}{Vaswani, A.} \emph{et~al.}
\newblock \bibinfo{journal}{\bibinfo{title}{Attention is all you need}}.
\newblock {\emph{\JournalTitle{Advances in neural information processing
  systems}}} \textbf{\bibinfo{volume}{30}} (\bibinfo{year}{2017}).

\bibitem{achiam2023gpt}
\bibinfo{author}{Achiam, J.} \emph{et~al.}
\newblock \bibinfo{journal}{\bibinfo{title}{Gpt-4 technical report}}.
\newblock {\emph{\JournalTitle{arXiv preprint arXiv:2303.08774}}}
  (\bibinfo{year}{2023}).

\bibitem{ramesh2021zero}
\bibinfo{author}{Ramesh, A.} \emph{et~al.}
\newblock \bibinfo{title}{Zero-shot text-to-image generation}.
\newblock In \emph{\bibinfo{booktitle}{International conference on machine
  learning}}, \bibinfo{pages}{8821--8831} (\bibinfo{organization}{Pmlr},
  \bibinfo{year}{2021}).

\bibitem{ramsauer2020hopfield}
\bibinfo{author}{Ramsauer, H.} \emph{et~al.}
\newblock \bibinfo{journal}{\bibinfo{title}{Hopfield networks is all you
  need}}.
\newblock {\emph{\JournalTitle{arXiv preprint arXiv:2008.02217}}}
  (\bibinfo{year}{2020}).

\bibitem{olsson2022context}
\bibinfo{author}{Olsson, C.} \emph{et~al.}
\newblock \bibinfo{journal}{\bibinfo{title}{In-context learning and induction
  heads}}.
\newblock {\emph{\JournalTitle{arXiv preprint arXiv:2209.11895}}}
  (\bibinfo{year}{2022}).

\bibitem{von2023transformers}
\bibinfo{author}{Von~Oswald, J.} \emph{et~al.}
\newblock \bibinfo{title}{Transformers learn in-context by gradient descent}.
\newblock In \emph{\bibinfo{booktitle}{International Conference on Machine
  Learning}}, \bibinfo{pages}{35151--35174} (\bibinfo{organization}{PMLR},
  \bibinfo{year}{2023}).

\bibitem{zhang2024trained}
\bibinfo{author}{Zhang, R.}, \bibinfo{author}{Frei, S.} \&
  \bibinfo{author}{Bartlett, P.~L.}
\newblock \bibinfo{journal}{\bibinfo{title}{Trained transformers learn linear
  models in-context}}.
\newblock {\emph{\JournalTitle{Journal of Machine Learning Research}}}
  \textbf{\bibinfo{volume}{25}}, \bibinfo{pages}{1--55} (\bibinfo{year}{2024}).

\bibitem{gu2021efficiently}
\bibinfo{author}{Gu, A.}, \bibinfo{author}{Goel, K.} \& \bibinfo{author}{Re,
  C.}
\newblock \bibinfo{title}{Efficiently modeling long sequences with structured
  state spaces}.
\newblock In \emph{\bibinfo{booktitle}{International Conference on Learning
  Representations}} (\bibinfo{year}{2021}).

\bibitem{orvieto2023resurrecting}
\bibinfo{author}{Orvieto, A.} \emph{et~al.}
\newblock \bibinfo{title}{Resurrecting recurrent neural networks for long
  sequences}.
\newblock In \emph{\bibinfo{booktitle}{International Conference on Machine
  Learning}}, \bibinfo{pages}{26670--26698} (\bibinfo{organization}{PMLR},
  \bibinfo{year}{2023}).

\bibitem{poirazi2020illuminating}
\bibinfo{author}{Poirazi, P.} \& \bibinfo{author}{Papoutsi, A.}
\newblock \bibinfo{journal}{\bibinfo{title}{Illuminating dendritic function
  with computational models}}.
\newblock {\emph{\JournalTitle{Nature Reviews Neuroscience}}}
  \textbf{\bibinfo{volume}{21}}, \bibinfo{pages}{303--321}
  (\bibinfo{year}{2020}).

\bibitem{larkum2013cellular}
\bibinfo{author}{Larkum, M.}
\newblock \bibinfo{journal}{\bibinfo{title}{A cellular mechanism for cortical
  associations: an organizing principle for the cerebral cortex}}.
\newblock {\emph{\JournalTitle{Trends in neurosciences}}}
  \textbf{\bibinfo{volume}{36}}, \bibinfo{pages}{141--151}
  (\bibinfo{year}{2013}).

\bibitem{urbanczik2014learning}
\bibinfo{author}{Urbanczik, R.} \& \bibinfo{author}{Senn, W.}
\newblock \bibinfo{journal}{\bibinfo{title}{Learning by the dendritic
  prediction of somatic spiking}}.
\newblock {\emph{\JournalTitle{Neuron}}} \textbf{\bibinfo{volume}{81}},
  \bibinfo{pages}{521--528} (\bibinfo{year}{2014}).

\bibitem{guerguiev2017towards}
\bibinfo{author}{Guerguiev, J.}, \bibinfo{author}{Lillicrap, T.~P.} \&
  \bibinfo{author}{Richards, B.~A.}
\newblock \bibinfo{journal}{\bibinfo{title}{Towards deep learning with
  segregated dendrites}}.
\newblock {\emph{\JournalTitle{eLife}}} \textbf{\bibinfo{volume}{6}},
  \bibinfo{pages}{e22901} (\bibinfo{year}{2017}).

\bibitem{sacramento2018dendritic}
\bibinfo{author}{Sacramento, J.}, \bibinfo{author}{Ponte~Costa, R.},
  \bibinfo{author}{Bengio, Y.} \& \bibinfo{author}{Senn, W.}
\newblock \bibinfo{title}{Dendritic cortical microcircuits approximate the
  backpropagation algorithm}.
\newblock In \bibinfo{editor}{Bengio, S.} \emph{et~al.} (eds.)
  \emph{\bibinfo{booktitle}{Advances in Neural Information Processing Systems
  31}}, \bibinfo{pages}{8721--8732} (\bibinfo{publisher}{Curran Associates,
  Inc.}, \bibinfo{year}{2018}).

\bibitem{payeur2021burst}
\bibinfo{author}{Payeur, A.}, \bibinfo{author}{Guerguiev, J.},
  \bibinfo{author}{Zenke, F.}, \bibinfo{author}{Richards, B.~A.} \&
  \bibinfo{author}{Naud, R.}
\newblock \bibinfo{journal}{\bibinfo{title}{Burst-dependent synaptic plasticity
  can coordinate learning in hierarchical circuits}}.
\newblock {\emph{\JournalTitle{Nature neuroscience}}}
  \textbf{\bibinfo{volume}{24}}, \bibinfo{pages}{1010--1019}
  (\bibinfo{year}{2021}).

\bibitem{capone2023beyond}
\bibinfo{author}{Capone, C.}, \bibinfo{author}{Lupo, C.},
  \bibinfo{author}{Muratore, P.} \& \bibinfo{author}{Paolucci, P.~S.}
\newblock \bibinfo{journal}{\bibinfo{title}{Beyond spiking networks: The
  computational advantages of dendritic amplification and input segregation}}.
\newblock {\emph{\JournalTitle{Proceedings of the National Academy of
  Sciences}}} \textbf{\bibinfo{volume}{120}}, \bibinfo{pages}{e2220743120}
  (\bibinfo{year}{2023}).

\bibitem{laskin2022context}
\bibinfo{author}{Laskin, M.} \emph{et~al.}
\newblock \bibinfo{journal}{\bibinfo{title}{In-context reinforcement learning
  with algorithm distillation}}.
\newblock {\emph{\JournalTitle{arXiv preprint arXiv:2210.14215}}}
  (\bibinfo{year}{2022}).

\bibitem{meulemans2022least}
\bibinfo{author}{Meulemans, A.}, \bibinfo{author}{Zucchet, N.},
  \bibinfo{author}{Kobayashi, S.}, \bibinfo{author}{Von~Oswald, J.} \&
  \bibinfo{author}{Sacramento, J.}
\newblock \bibinfo{journal}{\bibinfo{title}{The least-control principle for
  local learning at equilibrium}}.
\newblock {\emph{\JournalTitle{Advances in Neural Information Processing
  Systems}}} \textbf{\bibinfo{volume}{35}}, \bibinfo{pages}{33603--33617}
  (\bibinfo{year}{2022}).

\bibitem{feulner2022feedback}
\bibinfo{author}{Feulner, B.}, \bibinfo{author}{Perich, M.~G.},
  \bibinfo{author}{Miller, L.~E.}, \bibinfo{author}{Clopath, C.} \&
  \bibinfo{author}{Gallego, J.~A.}
\newblock \bibinfo{journal}{\bibinfo{title}{Feedback-based motor control can
  guide plasticity and drive rapid learning}}.
\newblock {\emph{\JournalTitle{bioRxiv}}} \bibinfo{pages}{2022--10}
  (\bibinfo{year}{2022}).

\bibitem{jiang2024dynamic}
\bibinfo{author}{Jiang, L.~P.} \& \bibinfo{author}{Rao, R.~P.}
\newblock \bibinfo{journal}{\bibinfo{title}{Dynamic predictive coding: A model
  of hierarchical sequence learning and prediction in the neocortex}}.
\newblock {\emph{\JournalTitle{PLoS Computational Biology}}}
  \textbf{\bibinfo{volume}{20}}, \bibinfo{pages}{e1011801}
  (\bibinfo{year}{2024}).

\bibitem{perez2018film}
\bibinfo{author}{Perez, E.}, \bibinfo{author}{Strub, F.},
  \bibinfo{author}{De~Vries, H.}, \bibinfo{author}{Dumoulin, V.} \&
  \bibinfo{author}{Courville, A.}
\newblock \bibinfo{title}{Film: Visual reasoning with a general conditioning
  layer}.
\newblock In \emph{\bibinfo{booktitle}{Proceedings of the AAAI conference on
  artificial intelligence}}, vol.~\bibinfo{volume}{32} (\bibinfo{year}{2018}).

\bibitem{wybo2023nmda}
\bibinfo{author}{Wybo, W.~A.} \emph{et~al.}
\newblock \bibinfo{journal}{\bibinfo{title}{Nmda-driven dendritic modulation
  enables multitask representation learning in hierarchical sensory processing
  pathways}}.
\newblock {\emph{\JournalTitle{Proceedings of the National Academy of
  Sciences}}} \textbf{\bibinfo{volume}{120}}, \bibinfo{pages}{e2300558120}
  (\bibinfo{year}{2023}).

\bibitem{de2022multiple}
\bibinfo{author}{De~Pitt{\`a}, M.} \& \bibinfo{author}{Brunel, N.}
\newblock \bibinfo{journal}{\bibinfo{title}{Multiple forms of working memory
  emerge from synapse--astrocyte interactions in a neuron--glia network
  model}}.
\newblock {\emph{\JournalTitle{Proceedings of the National Academy of
  Sciences}}} \textbf{\bibinfo{volume}{119}}, \bibinfo{pages}{e2207912119}
  (\bibinfo{year}{2022}).

\bibitem{berry1985bandit}
\bibinfo{author}{Berry, D.~A.} \& \bibinfo{author}{Fristedt, B.}
\newblock \bibinfo{journal}{\bibinfo{title}{Bandit problems: sequential
  allocation of experiments (monographs on statistics and applied
  probability)}}.
\newblock {\emph{\JournalTitle{London: Chapman and Hall}}}
  \textbf{\bibinfo{volume}{5}}, \bibinfo{pages}{7--7} (\bibinfo{year}{1985}).

\bibitem{morris1981spatial}
\bibinfo{author}{Morris, R.~G.}
\newblock \bibinfo{journal}{\bibinfo{title}{Spatial localization does not
  require the presence of local cues}}.
\newblock {\emph{\JournalTitle{Learning and motivation}}}
  \textbf{\bibinfo{volume}{12}}, \bibinfo{pages}{239--260}
  (\bibinfo{year}{1981}).

\bibitem{abraham2019plasticity}
\bibinfo{author}{Abraham, W.~C.}, \bibinfo{author}{Jones, O.~D.} \&
  \bibinfo{author}{Glanzman, D.~L.}
\newblock \bibinfo{journal}{\bibinfo{title}{Is plasticity of synapses the
  mechanism of long-term memory storage?}}
\newblock {\emph{\JournalTitle{NPJ science of learning}}}
  \textbf{\bibinfo{volume}{4}}, \bibinfo{pages}{9} (\bibinfo{year}{2019}).

\bibitem{schmidhuber1992learning}
\bibinfo{author}{Schmidhuber, J.}
\newblock \bibinfo{journal}{\bibinfo{title}{Learning to control fast-weight
  memories: An alternative to dynamic recurrent networks}}.
\newblock {\emph{\JournalTitle{Neural Computation}}}
  \textbf{\bibinfo{volume}{4}}, \bibinfo{pages}{131--139}
  (\bibinfo{year}{1992}).

\bibitem{irie2021going}
\bibinfo{author}{Irie, K.}, \bibinfo{author}{Schlag, I.},
  \bibinfo{author}{Csord{\'a}s, R.} \& \bibinfo{author}{Schmidhuber, J.}
\newblock \bibinfo{journal}{\bibinfo{title}{Going beyond linear transformers
  with recurrent fast weight programmers}}.
\newblock {\emph{\JournalTitle{Advances in neural information processing
  systems}}} \textbf{\bibinfo{volume}{34}}, \bibinfo{pages}{7703--7717}
  (\bibinfo{year}{2021}).

\bibitem{ba2016using}
\bibinfo{author}{Ba, J.}, \bibinfo{author}{Hinton, G.~E.},
  \bibinfo{author}{Mnih, V.}, \bibinfo{author}{Leibo, J.~Z.} \&
  \bibinfo{author}{Ionescu, C.}
\newblock \bibinfo{journal}{\bibinfo{title}{Using fast weights to attend to the
  recent past}}.
\newblock {\emph{\JournalTitle{Advances in neural information processing
  systems}}} \textbf{\bibinfo{volume}{29}} (\bibinfo{year}{2016}).

\bibitem{chalvidal2022meta}
\bibinfo{author}{Chalvidal, M.}, \bibinfo{author}{Serre, T.} \&
  \bibinfo{author}{VanRullen, R.}
\newblock \bibinfo{journal}{\bibinfo{title}{Meta-reinforcement learning with
  self-modifying networks}}.
\newblock {\emph{\JournalTitle{Advances in Neural Information Processing
  Systems}}} \textbf{\bibinfo{volume}{35}}, \bibinfo{pages}{7838--7851}
  (\bibinfo{year}{2022}).

\bibitem{miconi2018differentiable}
\bibinfo{author}{Miconi, T.}, \bibinfo{author}{Stanley, K.} \&
  \bibinfo{author}{Clune, J.}
\newblock \bibinfo{title}{Differentiable plasticity: training plastic neural
  networks with backpropagation}.
\newblock In \emph{\bibinfo{booktitle}{International Conference on Machine
  Learning}}, \bibinfo{pages}{3559--3568} (\bibinfo{organization}{PMLR},
  \bibinfo{year}{2018}).

\bibitem{zucchet2023gated}
\bibinfo{author}{Zucchet, N.} \emph{et~al.}
\newblock \bibinfo{journal}{\bibinfo{title}{Gated recurrent neural networks
  discover attention}}.
\newblock {\emph{\JournalTitle{arXiv preprint arXiv:2309.01775}}}
  (\bibinfo{year}{2023}).

\bibitem{wilson1972excitatory}
\bibinfo{author}{Wilson, H.~R.} \& \bibinfo{author}{Cowan, J.~D.}
\newblock \bibinfo{journal}{\bibinfo{title}{Excitatory and inhibitory
  interactions in localized populations of model neurons}}.
\newblock {\emph{\JournalTitle{Biophysical journal}}}
  \textbf{\bibinfo{volume}{12}}, \bibinfo{pages}{1--24} (\bibinfo{year}{1972}).

\bibitem{yildiz2012re}
\bibinfo{author}{Yildiz, I.~B.}, \bibinfo{author}{Jaeger, H.} \&
  \bibinfo{author}{Kiebel, S.~J.}
\newblock \bibinfo{journal}{\bibinfo{title}{Re-visiting the echo state
  property}}.
\newblock {\emph{\JournalTitle{Neural networks}}}
  \textbf{\bibinfo{volume}{35}}, \bibinfo{pages}{1--9} (\bibinfo{year}{2012}).

\bibitem{huang2006extreme}
\bibinfo{author}{Huang, G.-B.}, \bibinfo{author}{Zhu, Q.-Y.} \&
  \bibinfo{author}{Siew, C.-K.}
\newblock \bibinfo{journal}{\bibinfo{title}{Extreme learning machine: theory
  and applications}}.
\newblock {\emph{\JournalTitle{Neurocomputing}}} \textbf{\bibinfo{volume}{70}},
  \bibinfo{pages}{489--501} (\bibinfo{year}{2006}).

\bibitem{larkum2013}
\bibinfo{author}{Larkum, M.}
\newblock \bibinfo{journal}{\bibinfo{title}{A cellular mechanism for cortical
  associations: an organizing principle for the cerebral cortex}}.
\newblock {\emph{\JournalTitle{Trends in Neurosciences}}}
  \textbf{\bibinfo{volume}{36}}, \bibinfo{pages}{141 -- 151},
  \doiprefix\url{https://doi.org/10.1016/j.tins.2012.11.006}
  (\bibinfo{year}{2013}).

\bibitem{shai2015physiology}
\bibinfo{author}{Shai, A.~S.}, \bibinfo{author}{Anastassiou, C.~A.},
  \bibinfo{author}{Larkum, M.~E.} \& \bibinfo{author}{Koch, C.}
\newblock \bibinfo{journal}{\bibinfo{title}{Physiology of layer 5 pyramidal
  neurons in mouse primary visual cortex: coincidence detection through
  bursting}}.
\newblock {\emph{\JournalTitle{PLoS computational biology}}}
  \textbf{\bibinfo{volume}{11}}, \bibinfo{pages}{e1004090}
  (\bibinfo{year}{2015}).

\bibitem{bellec2020}
\bibinfo{author}{Bellec, G.} \emph{et~al.}
\newblock \bibinfo{journal}{\bibinfo{title}{A solution to the learning dilemma
  for recurrent networks of spiking neurons}}.
\newblock {\emph{\JournalTitle{Nature communications}}}
  \textbf{\bibinfo{volume}{11}}, \bibinfo{pages}{1--15} (\bibinfo{year}{2020}).

\bibitem{capone2022towards}
\bibinfo{author}{Capone, C.} \& \bibinfo{author}{Paolucci, P.~S.}
\newblock \bibinfo{journal}{\bibinfo{title}{Towards biologically plausible
  dreaming and planning}}.
\newblock {\emph{\JournalTitle{arXiv preprint arXiv:2205.10044}}}
  (\bibinfo{year}{2022}).

\bibitem{gu2023mamba}
\bibinfo{author}{Gu, A.} \& \bibinfo{author}{Dao, T.}
\newblock \bibinfo{journal}{\bibinfo{title}{Mamba: Linear-time sequence
  modeling with selective state spaces}}.
\newblock {\emph{\JournalTitle{arXiv preprint arXiv:2312.00752}}}
  (\bibinfo{year}{2023}).

\bibitem{hart2020embedding}
\bibinfo{author}{Hart, A.}, \bibinfo{author}{Hook, J.} \&
  \bibinfo{author}{Dawes, J.}
\newblock \bibinfo{journal}{\bibinfo{title}{Embedding and approximation
  theorems for echo state networks}}.
\newblock {\emph{\JournalTitle{Neural Networks}}}
  \textbf{\bibinfo{volume}{128}}, \bibinfo{pages}{234--247}
  (\bibinfo{year}{2020}).

\bibitem{courellis2024abstract}
\bibinfo{author}{Courellis, H.~S.} \emph{et~al.}
\newblock \bibinfo{journal}{\bibinfo{title}{Abstract representations emerge in
  human hippocampal neurons during inference}}.
\newblock {\emph{\JournalTitle{Nature}}} \bibinfo{pages}{1--9}
  (\bibinfo{year}{2024}).

\end{thebibliography}

\newpage

\appendix

\renewcommand{\thefigure}{A\arabic{figure}}
\setcounter{figure}{0}


\section*{Approximating gradients with gain modulated architectures and relationship with previous work}

The relationship between attention-based transformer architectures and in-context learning was first noted in \cite{von2023transformers} where it was shown through constructive proof, confirmed by experiments, that linear attention layers can implement a gradient descent update for linear regression in its forward pass.
Building upon this insight, subsequent work\cite{zucchet2023gated} showed that a linear gated RNN can implement the same mechanism, showing that a linear two-layer gated RNN can replicate a linear transformer. The work proposes an implementation that uses $O(d^2)$ hidden units and has $O(d^4)$ trainable weights, that can be reduced to $O(d^3)$ by side gating. 
Our gain-modulated architecture presents a generalized version of the linear gated RNNs used in the deep learning literature \cite{zucchet2023gated}. 
To see this, consider two linear \footnote{For simplicity, here we assume $\vb{b}^{ap}, \beta = 0,\ \gamma=1$ and consider a linear activation function $\phi = Id$.} GMNNs implementing the functions $\textcolor{orange}{Y_{\Theta^y}}(\vb{z}, \vb{w})$, $\textcolor{darkgreen}{W_{\Theta^w}}(\vb{z}, y^{targ} - y)$, dynamically performing a ICL linear regression task that requires to map temporal dependent features $z(t)\in \mathbb{R}^{N_{in}}$ to predict an output $\vb{y}^{targ}\in \mathbb{R}^{N_{out}}$:

\begin{equation}\label{eq:linear-rnn}
\begin{cases}
\vb{y} & =  \textcolor{orange}{Y_{\Theta^y}}(\vb{z}, \vb{w}) = \Theta^y \left(R^{ap}_y \vb{w}\right) \odot (R_y\vb{z}(t)) \\
\tau_w \, \dot{\vb{w}} & =  \textcolor{darkgreen}{W_{\Theta^w}}(\vb{z}, \vb{y}^{targ} - \vb{y}) = \Theta^w \left(R^{ap}_w\left(\hat{y} - y\right)\right) \odot (R_w\vb{z}(t))
\end{cases}
\end{equation} 
This architecture can be seen as a particular instance of the RNN with side gating (see Equation (15) in \cite{zucchet2023gated}). Interestingly, a similar mechanism is also present in the recently proposed Mamba layer \cite{gu2023mamba}.

In the rest of this appendix, we will demonstrate that an architecture with gain modulation is better suited to approximate the gradient terms involved in dynamical learning within a simple univariate linear regression task on the features $\vb{z}(t)$. In this task, as described in the Methods section, the functions $ Y $ and $ W $ are represented by a dot product and a scalar-vector product, respectively. To simplify the analysis, we employ two GMNNs (gain-modulated neural networks), modeled as extreme learning machines (ELMs), without recurrent connections to approximate these functions. While an architecture without recurrent connections suffices for this task, more complex scenarios involving temporal credit assignment require recurrent connections, as shown in \FigRef{figs2}. Notably, in the mathematical literature, the general approximation capability of an Echo State Network (ESN) can be rigorously understood through its relationship to an associated extreme learning machine \cite{hart2020embedding}.

\paragraph{Scalar product}
We first consider the task of approximating a scalar product between virtual weights $\vb{w}\in \mathbb{R}^{N_{in}}$ and features $\vb{z}\in\mathbb{R}^{N_{in}}$.
To achieve this, we train the readout weights $\Theta$ of a gain modulated network $\mathrm{GMNN}^\gamma_\Theta(\vb{z}|\vb{w})$ with $N_h$ hidden features to approximate the function $dot(\vb{x}, \vb{w}) = \sum_{i=1}^{N_{in}} z_iw_i$. 

\begin{figure*}[ht!]
\centering
\includegraphics[width=\linewidth]{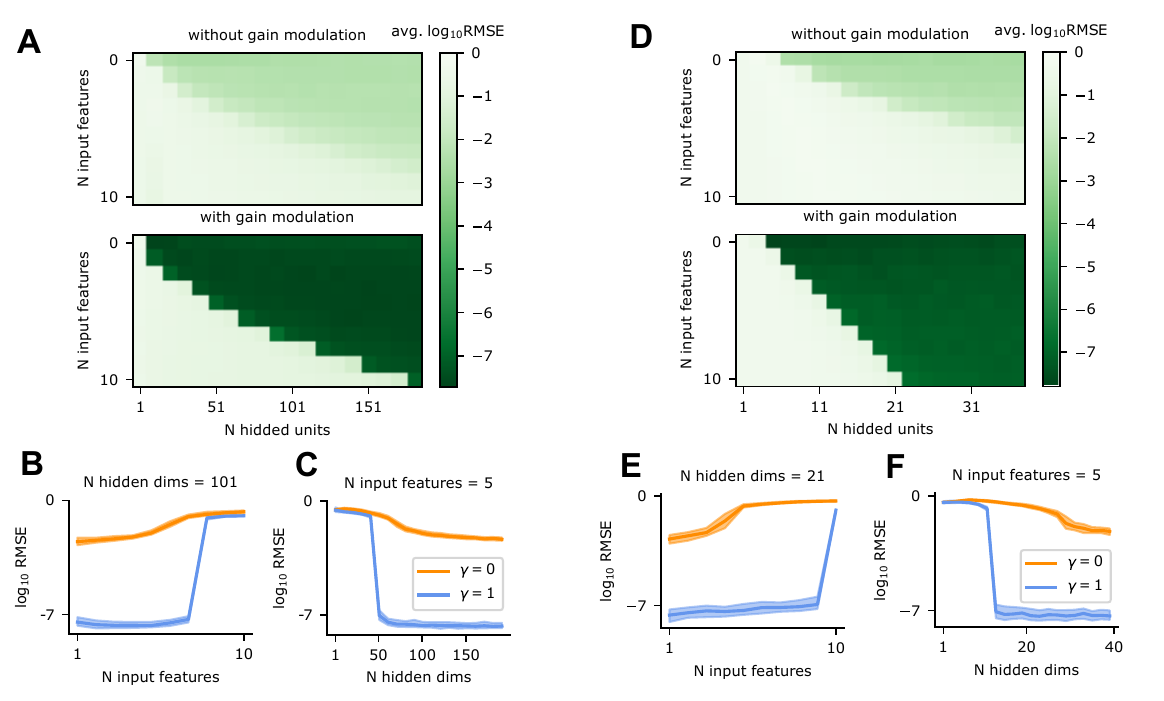}
\caption{\textbf{A} Test errors for the scalar product approximation task, varying the number of features $N_{in}$, hidden units $N_h$. We compare architectures with gain modulation ($\gamma=1$, bottom) with architectures without gain modulation ($\gamma=0$, bottom). 
\textbf{B} Test errors for the dot product approximation task, varying the number of features $N_{in}$, and fixing the number of hidden units $N_h=101$. Solid lines indicate the median over 100 trained models, while the filled region indicates the $20/80th$-percentile interval. \textbf{C} Same as \textbf{B} but fixing the number of features $N_{in}=5$, and varying the number of hidden units $N_h=101$. 
\textbf{D, E, F} Same as panels \textbf{A, B, C} but for the scalar-vector product approximation task.
}
\label{figs1}
\end{figure*}

We use a training dataset composed of $1000$ pairs $(\vb{z}, \vb{w})$ uniformly sampled in the hypercube $[0,1]^{2N_{in}}$. The test set consists of the same number of pairs sampled in the hypercube $[-1,1]^{2N_{in}}$.
We compare an architecture with gain modulation ($\gamma=1$) with an architecture without gain modulation $\gamma=0$. For each of the two settings, we select the best hyperparameters by fixing the dimensionality of the inputs ($N_{in}=5$) and hidden units ($N_h=101$) and varying the standard deviation of the projection matrices $R_{ij}, R^{ap}_{ij} \sim \mathcal{N}(0, \sigma_R^2),\ \log_{10}(\sigma_R)\in \{-2, -1.8, \cdots, 0\}$, the standard deviation of the bias $b_i\sim \mathcal{N}(0, \sigma_b^2),\ \sigma_b\in \{0, 0.1, \cdots, 1\}$ and the nonlinearity type $\phi \in \{\mathrm{tanh}, \mathrm{softplus}\}$. 
The best hyperparameters found were used in the models to test the approximation performance of the dot product varying the number of input features ($N_{in}\in [1, 10]\cap\mathbb{N}$) and hidden units ($N_h\in [1, 200]\cap\mathbb{N}$). 
Results are shown in \FigRef{figs1} A, B, C. We see that while both architectures require a quadratic number of hidden units to achieve low error on the test set, the gain-modulated network after a threshold number of units is hidden units is reached, is able to perfectly approximate the scalar product function consistently achieving $\sim 10^{-7}$ RMSE error. At the same time, the error in the network without gain modulation remains several orders of magnitude higher.

\paragraph{Scalar-vector product}
 We then tackle the task of approximating a vector-scalar product between features $\vb{x}\in \mathbb{R}^{N_{in}}$ and features and error $\vb{e}\in\mathbb{R}$.
To achieve this, we train the readout weights $\Theta$ of a gain modulated network $\mathrm{GMNN}^\gamma_\Theta(\vb{x}|e)$ with $N_h$ hidden features to approximate the function $prod(\vb{x}, e) = e\vb{x}\in\mathbb{R}^{N_{in}}$. 

We used a training dataset is composed of $1000$ pairs $(\vb{x}, e)$ uniformly sampled in the hypercube $[0,1]^{N_{in}}\times[0,1]$. The test set consists of the same number of pairs sampled in the hypercube $[-1,1]^{N_{in}}\times[-1,1]$.

We compare an architecture with gain modulation ($\gamma=1$) with an architecture without gain modulation $\gamma=0$. For each of the two settings, we select the best hyperparameters by fixing the dimensionality of the inputs ($N_{in}=5$) and hidden units ($N_h=21$) and varying the standard deviation of the projection matrices 
$R_{ij}\sim \mathcal{N}(0, \sigma_R^2/N_{in}), R^{ap}_{ij}\sim \mathcal{N}(0, \sigma_R^2, \ \log_{10}(\sigma_R)\in \{-2, -1.8, \cdots, 0\}$, the standard deviation of the bias $b_i\sim \mathcal{N}(0, \sigma_b^2),\ \sigma_b\in \{0, 0.1, \cdots, 1\}$ and the nonlinearity type $\phi \in \{\mathrm{tanh}, \mathrm{softplus}\}$. 
The best hyperparameters found were used in the models to test the approximation performance of the dot product varying the number of input features ($N_{in}\in [1, 10]\cap\mathbb{N}$) and hidden units ($N_h\in [1, 40]\cap\mathbb{N}$). 
Results are shown in \FigRef{figs1} D, E, F. We see that the network without gain modulation requires a quadratic number of hidden units to achieve low error on the test set. As in the scalar product case, the gain-modulated network can nearly perfectly approximate the target function after a threshold number of hidden units is reached. Significantly, in this case, the threshold scales linearly with $\mathbb{R}^{N_{in}}$. 

\paragraph{On the number of hidden units and trainable weights needed in our model}
In general, for gain-modulated architectures, we experimentally found a linear scaling for the W network and a quadratic scaling for Y (as a function of the number $N_{in}$ of independent inputs). The scaling for Y can be further reduced to linear by using diagonal (or block diagonal) projection matrices R and $R^{ap}$. 
The empirically found scaling can be easily understood assuming the gain-modulated network operates in a near linear regime (which here is sufficient, since we are learning a linear relationship between the features and the readout, this is confirmed by the low input variance found by the hyperparameter search). Considering a first-order Taylor expansion, the hidden units extract features that are (random) linear combinations of products of apical and basal inputs. Since the target function (both for Y and W) is also a linear combination products of apical and basal inputs, it is sufficient for the readout to linearly recombine these terms into the right form.
Consider for example the network W. Each hidden unit j in the linear regime extracts a feature of the form:
$\sum_{k=1}^N A_{jk} \vb{z}_k + \sum_{k=1}^N C_{jk}\vb{z}_k e + \vb{b}_j\cdot e + \vb{d}_j$.
In matrix form the extracted features can be written as $A\vb{z} + C(e\vb{z}) + \vb{b}e + \vb{d}$, where $A$, $C$, $\vb{b}$, $\vb{d}$ are random matrix and vector coefficients determined by the random projection matrices and by the nonlinearity slope. 
The readout $\Theta$ needs to satisfy $\Theta(Az + C(ez) + \vb{b}e + \vb{d}) = e\vb{z}$. These are 2+2N linear constraints on $\Theta$ that can be satisfied only if the number of hidden units is $\ge 2+2N$. 
Notice that this is exactly the same scaling found experimentally (\FigRef{figs1}, panel D). 
The same considerations can be applied to the network Y.

Networks without gain modulation have significantly worse performance. To have products appear in the Taylor expansion for networks without gain modulation, we need a second-order Taylor expansion which produces a quadratic number of terms involving products between all input variables. This explains the quadratic scaling found for architectures without gain modulation in panels A and D of Fig. A1. 

Consider now the architecture in \EqRef{eq:linear-rnn} performing an online univariate linear regression task  ($N_{out}=1$): 

\begin{equation}\begin{cases}
y & =  \textcolor{orange}{Y_{\Theta^y}}(\vb{z}, \vb{w}) = \Theta^y \left(R^{ap}_y \vb{w}\right) \odot (R_y\vb{z}(t)) \\
\tau_w \, \dot{\vb{w}} & =  \textcolor{darkgreen}{W_{\Theta^w}}(\vb{z}, y^{targ} - y) = \Theta^w \left(R^{ap}_w\left(\hat{y} - y\right)\right) \odot (R_w\vb{z}(t))
\end{cases}
\end{equation}

Where $R^{ap}_y, R_y, R^{ap}_w, R_w$ are fixed random matrices respectively of dimension $N_y\times N_{in}, N_y\times N_{in}, N_w\times1, N_w\times N_{in}$, and $\Theta^y, \Theta^w$ are trainable readout matrices of sizes $1\times N_y, N_{in}\times N_w$.  From the previous experiment and theoretical analysis, we need $N_w\sim O(N_{in}^2)$ hidden units and $O(N_{in}^2)$ readout parameters for the $Y$ network. This number can be further reduced to $O(N_{in})$ both for the hidden units and trainable weights \footnote{This can be done assuming that $R^{ap}, R_w$ are diagonal matrices (and thus reducing the problem to estimating $N_{in}$ $1\times 1$ products)}. As for the network  $\textcolor{darkgreen}{W_{\Theta^w}}(\vb{z}, y^{targ} - y)$ approximating vector-scalar products, $N_y \sim O(N_{in})$ hidden units and $O(N_{in}^2)$ readout parameters are needed. 

For a multivariate $\mathbb{R}^{N_{in}} \to \mathbb{R}^{N_{out}}$ linear regression task, we can consider $N_{out}$ independent modules of the type described before. In this case, the number of trainable parameters is $O(N^2_{in}\cdot N_{out})$ with $O(N_{in}\cdot N_{out})$ hidden units ($O(N_{in}^2\cdot N_{out})$ in the more biologically plausible case).

\section*{Dynamical adaptation replaces synaptic plasticity: virtual weights can implement rule switching}

Consider a task inspired by the Wisconsin card-sorting test. Each card in a deck is defined by two features: color and shape. In each iteration, the agent receives a cue card and two guide cards—one matching the cue by color and the other by shape. The agent must select the correct guide card to receive a positive reward $R(t) = 1$. Initially, the correct choice is based on color matching. However, after 10 trials, the classification rule changes, rewarding shape matching instead. A successful agent must quickly adapt its decision-making to these contextual shifts.

 We consider a simple neural network that predicts reward probability for each (cue card, guide card) pair. Two populations, $ z_1 $ and $ z_2 $, encode similarity based on color and shape, respectively. A third population, $ y = \phi(w_1 z_1 + w_2 z_2)$, integrates this information to predict the reward probability. Here $ \phi(x) = \text{sigmoid}(x - \text{bias}) $ represents the neural transfer function. The weights $ w_1 $ and $ w_2 $ combine the features to determine the prediction, and the guide card with the highest predicted reward probability is selected. The agent dynamically updates the model based on received rewards to adapt to changing contexts.

Traditionally, weights $ w_1 $ and $ w_2 $ are considered synaptic weights, and learning is attributed to synaptic plasticity.
Alternatively, $ w_1 $ and $ w_2 $ can be reinterpreted as the activity of two additional populations acting as gain modulators
In our model, we refer to them as "virtual weights," as they replace traditional physical synaptic weights, effectively virtualizing the learning process. These modulatory populations dynamically reconfigure the responses of the network, enabling fast, flexible adaptations without the need for synaptic changes.
One of the key advantages of this virtualization is that it allows for learning rules that are not constrained by the strict locality requirements inherent to synaptic plasticity.

\begin{figure*}[ht!]
\centering
\includegraphics[width=\linewidth]{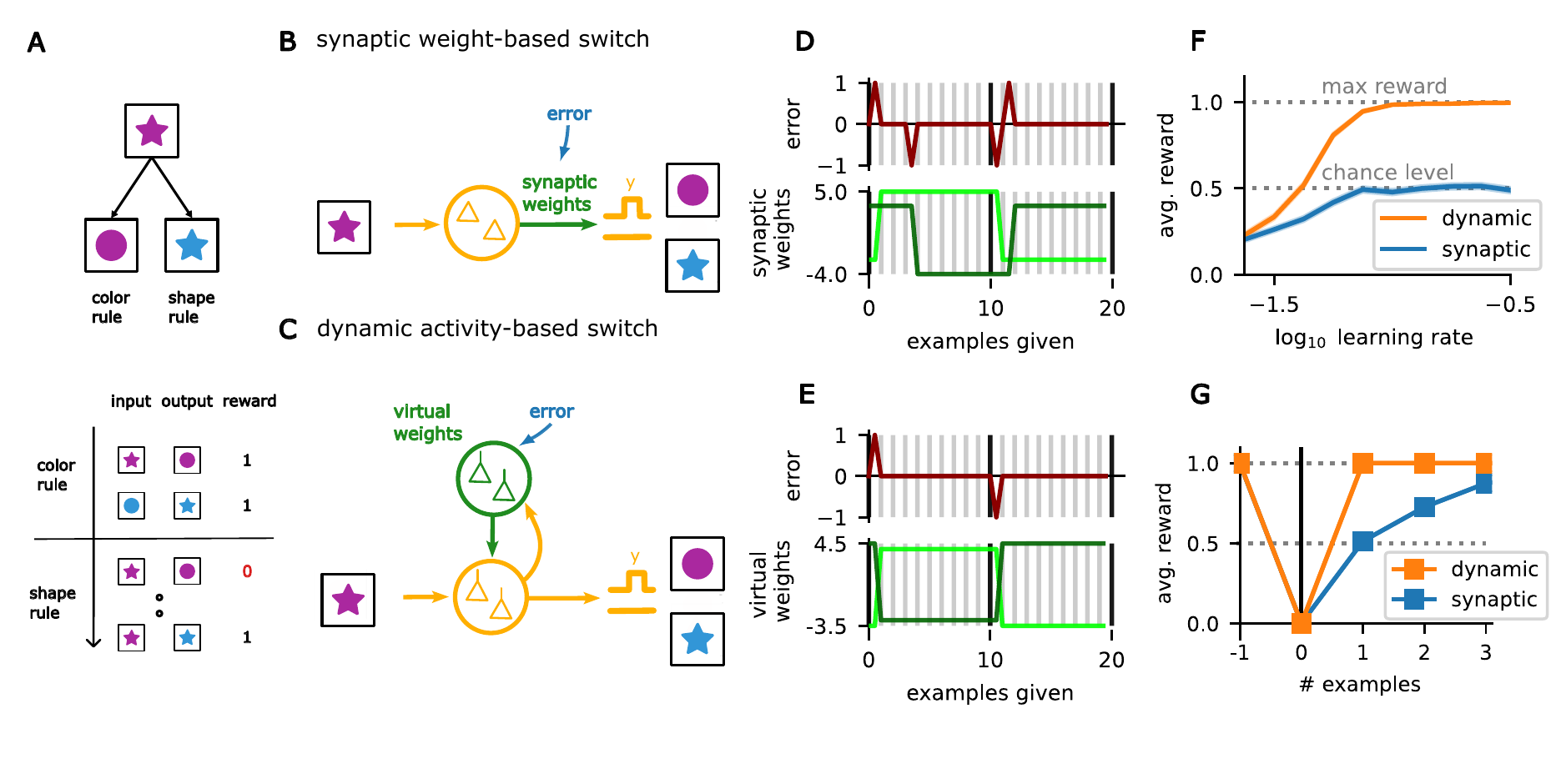}
\caption{\textbf{Adapting behavior in a context-dependent classification task.} \textbf{A.} Overview of the task structure: At each iteration, the agent is presented with a cue card and two guide cards, one matching the cue by color and the other by shape. The agent must select the correct guide card to receive a positive reward. The classification rule switches after 20 trials. \textbf{B.} Schematic representation of the proposed network architecture: synaptic plasticity of physical weights is replaced by dynamic adaptation of virtual weights \textbf{C.} that modulate the network response. \textbf{D.} Example temporal trajectory for the plastic network. Two in-context errors are needed to learn the correct rule. \textbf{E.} Example temporal trajectory for the network with dynamic adaptation: one in-context errors are needed to learn the correct rule. 
\textbf{F.} Average reward after 1 in-context example as a function of the learning rate $\eta$. Synaptic plasticity with a local delta rule does not go beyond the chance level. Dynamic adaptation with a non-local learning rule can achieve a perfect score.
The curve is obtained by averaging over 10 independent sessions of 2000 iterations each. The shaded region represents the standard deviation confidence interval.
\textbf{G.} Psychometric curve comparing average reward as a function of in-context examples. At $-1$ we report the average reward before the rule switch. The curve is obtained by averaging over 10 independent sessions of 2000 iterations each.
}
\label{fig0}
\end{figure*}

To support this point, in \FigRef{fig0} we compare synaptic learning using the gradient-based, delta rule $\Delta\vb{w} = \eta e \cdot \vb{z}$ with a non-local learning rule that force constraints $w_1(t) + w_2(t) = const$. This results in the update rule $\Delta w_j = \eta(-1)^j e\cdot(z_2 - z_1)$. Here $e(t) = (R(t) - y(t))$ represents the reward prediction error for the last choice and $\eta$ is the learning rate, controlling the speed of adaptation. As the latter learning rule features non-local terms, it can easily be implemented by a network that dynamically adjusts the activities of two modulatory populations of virtual weights $w_1, w_2$ without relying on synaptic updates (see \FigRef{fig0} C). 
In \FigRef{fig0} we compare the performance of an agent that learns with synaptic plasticity with an agent that performs dynamic adaptation. As we can see in panel $\ref{fig0}$ F, decreasing the timescales of the learning process, the network with dynamic adaptation achieves maximum reward after seeing only one in-context example, while the performance of the network with synaptic plasticity plateaus at near chance level. This is supported by the psychometric curve shown in \FigRef{fig0} G. Interestingly, the comparison between the two curves in \FigRef{fig0} G closely mirrors findings from a recent study \cite{courellis2024abstract}. In that study, some human participants displayed the ability to adjust their behavior after just one example, while others, with impaired cognitive abilities, had to relearn the previous classification rule through trial and error from scratch.

\section*{Additional details on the "dynamical learning of a temporal trajectory" experiment}

In the "dynamical learning of a temporal trajectory" experiment we test our architecture and non-synaptic learning approach on temporal tasks. The primary goal is to analyze the network's ability to generalize beyond the target frequencies used to pre-train our networks. Three networks are defined in this experiment: one uses a reservoir to compute temporal features and encode the target temporal trajectory, one predicts the required weight updates, and one predicts the scalar product, as discussed in the main text.

For clarity here we report the entire experiment protocol divided in the two different phases:
\begin{itemize}
    \item {\bf Phase 1: Development of the network structure (algorithm distillation)}
    \begin{itemize}
        \item We simulate the learning dynamics by integrating equation \eqref{eq:dyn-deltarule} with a suitable discretization timestep $dt$. We collect the trajectories $\{(\vb{z}_\alpha(t), \vb{w}_\alpha(t), e_\alpha(t))| \alpha \in \mathcal{A}_{ID}, t\in\{0, dt, \cdots, T\}\}$. 
        \item We tune the readout weights $\Theta_y$, $\Theta_w$ of the two GMRNN networks $Y_{\Theta_y}$, $W_{\Theta_w}$ to minimize ,respectively, the objectives $L_Y(\Theta_y)=\sum_{\alpha, t} \big|Y_{\theta_y}(\{\vb{z}_\alpha(t), \vb{w}_\alpha(t)\}) - \vb{w}_\alpha(t)\cdot \vb{z}_\alpha(t)\big|^2$ and $L_W(\Theta_w) = \sum_{\alpha, t} \big|W_{\Theta_w}(\{\vb{z}_\alpha(t), e_\alpha(t)\}) - \frac{d\vb{w}_i(t)}{dt}\big|^2$. This is done through a pseudo-inverse computation. 
    \end{itemize}
    \item {\bf Phase 2: Dynamical learning}
    \begin{itemize}
        \item {(\bf Open Loop)} The readout of the reservoir network is replaced with Y. All the synaptic weights are fixed.
        Given a novel sequence $y^{targ}_\alpha(t)$. Evolve neural dynamics following equation \eqref{eq:dyn-gd}.
        The neural activity $\vb{w}_\alpha(t)$ coding for the virtual weights is driven by $W_{\Theta_w}$. 
        \item {(\bf Closed Loop)} The external input is removed, the neural activity $\vb{w}_\alpha(t)$ coding for the virtual weights is frozen and $y(t) = Y_{\Theta_y}(z_\alpha(t), w_\alpha(t))$ is fed back as input 
    \end{itemize}

\end{itemize}

\begin{table}
  \caption{Simulation Parameters}
  \label{tab:parameters_exp2}
  \centering
  \begin{tabular}{lllll}
    \toprule
    Parameter & Symbol & Reservoir &  Gradient Net & Scalar Net \\
    \midrule
    Network Size & $N$ & 20 & 500&500  \\
    Input Dimension & $I$ & 1 & 100 + 10 & 100+100 \\
    Apical Input Dimension & $I^{ap}$ & 0 & 1 & 100 \\
    Output Dimension & $O$ & 1 & 100 & 1  \\
    Time Step & $dt$ & 0.005 \\
    Reservoir Time Constant & $\tau_{m_f}$ & 10 $dt$ &1 $dt$ &1 $dt$ \\
    Input weights var & $\sigma_{input}$ & 0.06& 0.06& 0.06 \\
    Apical Input weights var & $\sigma_{input}^{ap}$ & 0.0& 0.1& 0.1 \\
    Recurrent weights & $\sigma_{rec}$ & 0.99 / $\sqrt{N}$& 0.5 / $\sqrt{N}$& 0. / $\sqrt{N}$ \\
    Gain-modulation factor & $\gamma_{net}$ & 0.& 1.& 1. \\
    \bottomrule
  \end{tabular}
\end{table}

\paragraph{Training parameters}
We used for training five target trajectories $y^{targ}(t) = 0.8 sin(\omega_{targ} t)$, with five angular velocities, $\omega_{targ}$, ranging from $0.04$ to $0.08$.
The parameters used in the simulation are summarized in Table~\ref{tab:parameters_exp2}. Inputs are projected to the network through Gaussian weights with zero mean and variance $\sigma_{in}^2$ to distribute the input information across multiple units in the reservoir.

\paragraph{Loss function}

The online learning dynamics (equation \eqref{eq:dyn-deltarule}) for the readout weights $\vb{w}$ that the networks need to replicate is given by online gradient descent over the instantaneous reconstruction error:
\[
L(\vb{w},t) = | y(t) - y^{targ}(t)|^2 \quad \xrightarrow{} \quad \dot{\vb{w}}_j = -\eta\frac{\partial L}{\partial \vb{w}_j} = -\eta \left( y(t) - y^{\text{target}}(t) \right) \vb{z}_j(t),
\]
where $\eta$ is the learning rate.

\paragraph{Evaluation}
Performances are evaluated by measuring the MSE between the target trajectory ant the predicted one $y(t)$, for different values of the trajectory angular velocity, equally distribute between 0$.01$ and $0.1$.

\section*{Additional details on the reinforcement learning experiments}

For clarity here we report the entire experiment protocol divided into the two different phases:
\begin{itemize}
    \item {\bf Phase 1: Development of the network structure (algorithm distillation)}
    \begin{itemize}
        \item For each game $i \in [N_{games}]$ we select a random environment $\alpha_i$ uniformly form the ID set $\alpha_i \sim Unif(\mathcal{A}_{ID})$. 
        We then simulate an agent playing the game with policy $\vb{\pi} = \mathrm{Softmax}\left(Y(\vb{z}, \vb{w})\right)$, coupled with a policy gradient learning dynamics for the weights $\vb{w}$ (equation \eqref{eq:pg-nodisc} or \eqref{rl_discount} with a timestep $dt$) until the termination condition is reached for that game at time $T_i$.
        We collect the trajectory $\{(\vb{z}_{i}(t), \vb{w}_{i}(t), \vb{\pi}_{i}(t), r_{i}(t), a_{i}(t))|i \in [N_{games}], t\in\{0, dt, \cdots, T_i\}\}$. 
        \item We tune the readout weights $\Theta_y$, $\Theta_w$ of the two GMRNN $Y_{\Theta_y}$, $W_{\Theta_w}$ to minimize ,respectively, the objectives $L_Y(\Theta_y)=\sum_{\alpha, t} \big|Y_{\theta_y}(\{\vb{z}_\alpha(t), \vb{w}_\alpha(t)\}) - \vb{w}_\alpha(t)\cdot \vb{z}_\alpha(t)\big|^2$ and $L_W(\Theta_w) = \sum_{i, t} \big|W_{\Theta_w}(\{ r_{i}(t), a_{i}(t), \vb{\pi}_{i}(t), \vb{z}_{i}(t)\}) - \frac{d\vb{w}_i(t)}{dt}\big|^2$. This is done through a pseudo-inverse computation. 
    \end{itemize}
    \item {\bf Phase 2: Dynamical learning}
    \begin{itemize}
        \item The agent policy is now parametrized by the network $Y_{\Theta_y}$, while its internal learning dynamics is given by the network $W_{\Theta_w}$.  We now extract random environments $\alpha_i$ uniformly form the OOD set $\alpha_i \sim Unif(\mathcal{A}_{OOD})$.
        For each trial, we simulate the environment dynamics coupled with the network learning dynamics in equation \eqref{eq:dyn-pg}. 
        \item {(\bf evaluation)} The virtual weights are freezed. The agent parametrized by the network $Y_{\Theta_y}(\cdot, \vb{w})$ is now tested on all environments $\mathcal{A}$. 
    \end{itemize}

\end{itemize}

\subsection*{Reinforcement Learning and the Discount Factor}

\label{discount_factor_appendix}

Reinforcement Learning (RL) is a machine learning paradigm where an agent learns to make decisions by interacting with an environment to maximize cumulative rewards. The discount factor, usually denoted by $\gamma_d$ (where $0 \leq \gamma_d \leq 1$), is crucial in RL as it determines the importance of future rewards. The cumulative reward $R_t$ at time step $t$ is given by:

\[
R_t = r_t + \gamma_d r_{t+1} + \gamma^2_d r_{t+2} + \gamma^3_d r_{t+3} + \ldots
\]

In our experiment, the policy gradient update rule in the presence of the discount factor could be approximated as (see \cite{bellec2020}):

\begin{equation}
\begin{cases}
\dot{\vb{e}} & = -\frac{\vb{{e}}}{\tau_e} +  \left(\vb{\mathds{1}}_a - \vb{\pi} \right) \odot \vb{x} \\
\vb{y} & =  \vb{w} \cdot \vb{x} \\
\tau_w \, \dot{\vb{w}} & =  r \vb{e}\, 
\end{cases}
\label{rl_discount}
\end{equation}

where $\vb{e}$ is an exponential temporal filtering of the term $\left(\vb{\mathds{1}}_a - \vb{\pi} \right) \odot \vb{x} $, with a timescale $\tau_e = -\frac{1}{\log(\gamma_d)}$.
The importance of the discount factor lies in its ability to balance immediate and future rewards. A higher $\gamma_d$ values future rewards more significantly, encouraging the agent to consider the long-term consequences of its actions. Conversely, a lower $\gamma_d$ makes the agent prioritize immediate rewards.
This balance is essential for the stability of the learning process. An appropriately chosen $\gamma_d$ ensures stable learning and convergence; if $\gamma_d$ is too high, the agent might overvalue distant future rewards, leading to instability, while a very low $\gamma_d$ can result in short-sighted behavior.
In the limit $\gamma_d \to 0$ we recover Equation \eqref{eq:pg-nodisc} used in the main text. If the discount factor is zero, the agent's policy would change only when it is very close to the moment the reward is received. In a reaching task, the agent will only change its policy when it is very close to the reward, completely ignoring the need for long-term planning and making it unlikely to ever reach the goal from distant starting points.

\begin{figure*}[ht!]
\centering
\includegraphics[width=\linewidth]{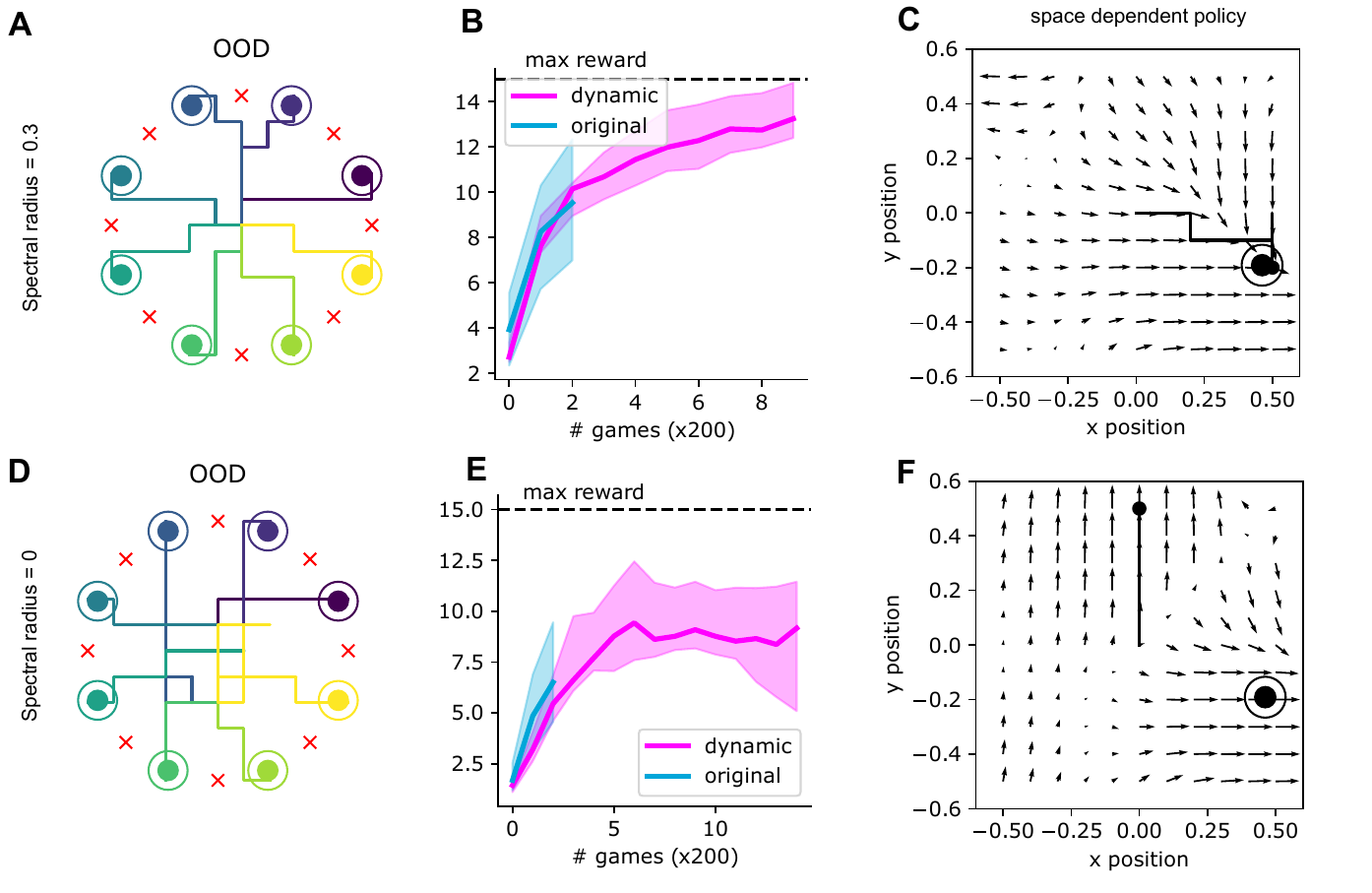}
\caption{ \textbf{Discount Factor and eligibility traces:}
\textbf{A.} Sample trajectories of the agent at the end of the learning procedure, for various target locations (color-coded), different from the ones used to train the gradient and the scalar product networks (red crosses).
\textbf{B.} Reward as a function of number of trials (or games), blue: policy gradient, pink: dynamical learning for new positions (OOD).
Line: median, shading: 20-th/80-th percentile range.
\textbf{C.} Arrows: policy at the end of the training as a function of the position. Line: sample trajectory of the agent at the end of learning. Small circle: agent position, double circle: target position.  
\textbf{D-E-F} same as in \textbf{A-B-C}, but the recurrent connections of the gradient network are set to zero.
}
\label{figs2}
\end{figure*}

To demonstrate that our approach performs the temporal computation required to consider future rewards, we compare the performances of our network (see \FigRef{figs2}.A-B) to a network without recurrent connections, which therefore cannot perform temporal computations. Recurrent connections enable a network to maintain a memory of past states and actions, effectively allowing it to use information from previous time steps to inform current decisions. Without these connections, a network operates purely on the current input without any contextual information from prior steps, thus lacking the ability to perform temporal computations (see \FigRef{figs2}.D-E).

Performances are higher in the first case. This can be visualized by looking at the policy at the end of the dynamic reinforcement learning for a specific target location. When the recurrent weights are set to zero, the policy points towards the target position only when nearby the target itself (see \FigRef{figs2}F), resulting in failure when the agent randomly moves in the wrong direction at the beginning of the task (see \FigRef{figs2}F, black line).

On the other hand, in the presence of recurrent weights, the proper policy (pointing towards the target, see \FigRef{figs2}C) is known even when far from the target, allowing optimal long-term planning (see \FigRef{figs2}C, black line).

\subsection*{Additional details on the bandit experiment}
\paragraph{Family of tasks}
We consider a family of tasks $\mathcal{A} = \{ID, OOD\}$, such that every element $\alpha \in \mathcal{A}$ represents a Bernoulli $K$-armed bandit problem with rewards $\{\hat r_i^{\alpha}\}_{i=1}^K$.
Each task $\alpha$ is specified by a set $P_\alpha \subset \{1\cdot, K\}$ of positive arms, such that 
\begin{align}
    Prob_\alpha({r|i}) = \hat{r}_i^{\alpha} = \begin{cases}
    {p} & i\in {P_\alpha} \\ 
    {1-p} & i\not\in {P_\alpha}
\end{cases}
\end{align}

In the main text experiment, we used $K=10$ and $P_{ID} := \{i \in [K]| i \equiv 0 (\text{mod} 2) \}$ and $P_{OOD} := \{i \in [K]| i \equiv 1 (\text{mod} 2) \}$.

\paragraph{Network details and hyperparameter search}
We compare an architecture with gain modulation ($\gamma=1$) with an architecture without gain modulation $\gamma=0$. For each of the two settings, we select the best hyperparameters by fixing the number of hidden units ($N_h=100$) and varying the standard deviation of the projection matrices $R_{ij} \sim \mathcal{N}(0, \sigma_R^2/20)\ \ R^{ap}_{ij}\sim \mathcal{N}(0, \sigma_R^2), \ \log_{10}(\sigma_R)\in \{-2, -1.8, \cdots, 0\}$, the standard deviation of the bias $b_i\sim \mathcal{N}(0, \sigma_b^2),\ \sigma_b\in \{0, 0.1, \cdots, 1\}$ and the nonlinearity type $\phi \in \{\mathrm{tanh}, \mathrm{softplus}\}$. For each point in the grid, we train 20 models using
The best hyperparameters found were used in the subsequent experiments 
In \FigRef{fig3} B, D the reported regrets are on models trained with $N_h=100$ hidden features

\paragraph{Regret per round}
Given an agent that plays in an environment $\alpha$ receiving rewards $\{r^{\alpha}(t)\}_{t\in\mathbb{N}_+}$, the regret per round $\rho(T)$ achieved by the agent at round $T$ is defined as: 
\begin{align}
    \rho(T) = \mu^{\alpha}_\star - \frac{1}{T}\sum_{t=1}^T r^\alpha(t)
\end{align}
Where $\mu^\alpha_*:= \max\{ \hat r_k^\alpha| k\in [K]\} = \max\{p, 1-p\} $ is the maximum expected reward that can be obtained in each round playing the optimal policy that selects one of the best arms with probability $1$.

\subsection*{Additional details on the darkroom experiment}

This experiment investigates the capability of out dynamical reinforcement learning approach, to improve an agent's ability to navigate a 2D environment. The agent is trained to locate a randomly placed, not observable food source, with its policy evolving over multiple episodes through.

The environment is a 2D grid where an agent starts at the center, aiming to reach a randomly positioned food item. The food positions vary every 600 episodes, introducing. The agent's movements are restricted within the grid, with actions limited to moving left, right, up, or down.

To represent the agent's position, a place cell encoder converts the agent's $(x, y)$ coordinates into a higher-dimensional feature vector using Gaussian functions. This encoding helps in effectively capturing spatial information.

The agent's policy is linear, with action probabilities derived from a softmax function applied to the encoded state. For this reason, only two networks are used for the task, the gradient and the scalar-product network see Table~\ref{tab:parameters_exp3} for details on parameters.

\paragraph{Data collection and pre-training}
The policy is updated using the policy gradient method, adjusting weights based on received rewards to improve decision-making over time.

Learning data are collected for 8 different positions of the food, equally distributed on a circle.

Two types of networks are used: one to model the agent's state dynamics and another to handle gradient updates necessary for learning. These networks influence the agent's internal state and learning process, enabling  policy refinement.

The two networks readout are trained with a linear regression on their readout weight, to reproduce the proper weight updates an scalar products.

\paragraph{Dynamical reinforcement learning and evaluation}

The agent is tested again on the 8 positions used for pre-training, and on 8 new positions.
Similarly to previous experiments, policy gradient is no longer used and replaced by virtual weights updated estimated by the gradient network, and used by the scalar-product network to predict agent policy.

Performance is measured by the total reward accumulated over episodes. Visualizing the agent's trajectories reveals its navigation efficiency and decision-making process.

\begin{table}
  \caption{Simulation Parameters}
  \label{tab:parameters_exp3}
  \centering
  \begin{tabular}{lllll}
    \toprule
    Parameter & Symbol &   Gradient Net & Scalar Net \\
    \midrule
    Network Size & $N$ &  500 & 1000  \\
    Input Dimension & $I$ & 25+4+10 & 4 $\times$ 25+4 $\times$ 25 \\
    Apical Input Dimension & $I^{ap}$ & 1 & 4 $\times$ 25 \\
    Output Dimension & $O$ & 4 $\times$ 25 & 4  \\
    Time Step & $dt$ & 0.005 \\
    Reservoir Time Constant & $\tau_{m_f}$ &1 $dt$ &1 $dt$ \\
    Input weights var & $\sigma_{input}$ & 0.01 & 0.01 \\
    Apical Input weights var & $\sigma_{input}$ & 0.1 & 0.1 \\
    Recurrent weights & $\sigma_{rec}$ & 0.5 / $\sqrt{N}$& 0. / $\sqrt{N}$ \\
    Gain-modulation factor & $\gamma_{net}$ & 1.& 1. \\
    \bottomrule
  \end{tabular}
\end{table}

\section*{Estimating the occurrence of dynamic adaptation}
To evaluate the ability of our biologically plausible recurrent network to adapt without synaptic plasticity, we tested it in a context-dependent locomotion task within the MuJoCo simulation environment. The task requires an Ant agent to reach a target location positioned either to the left or right at the end of a horizontal corridor. At the beginning of each trial, the ant does not have prior explicit information about the target direction and must infer it dynamically as the task unfolds.

\begin{figure*}[ht!]
\centering
\includegraphics[width=.7\linewidth]{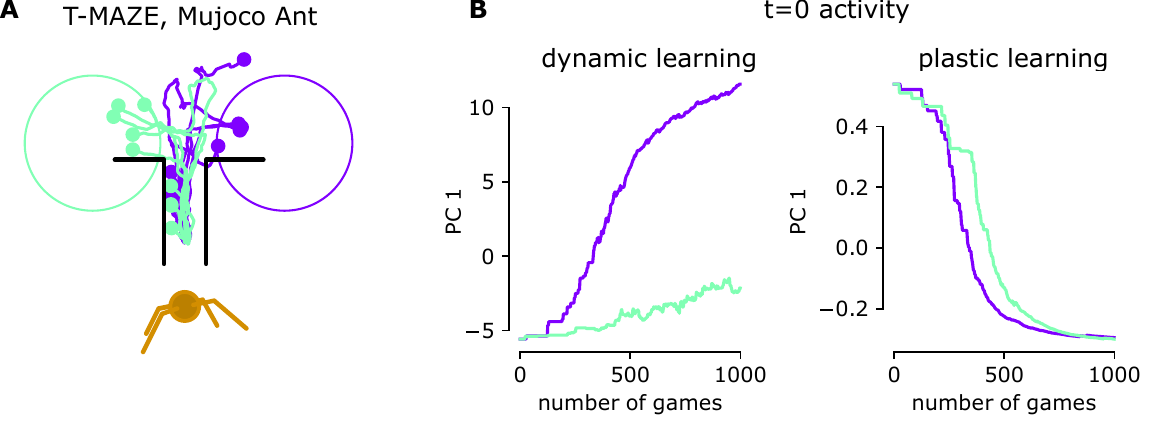}
\caption{\textbf{T-Maze task in Mujoco-ANT environment.} 
\textbf{A.} Task description, the ant has to reach a target location after an horizontal corridor.
\textbf{B.} Decoding of contextual information from the network activity at the initial time of each trial/game, as a function of the trial/game number in the case of dynamic and plastic learning (left and right respectively).
} 
\label{figs5}
\end{figure*}

\subsection*{Decoding Contextual Information from Network Activity}
To understand how contextual information is processed, we analyzed the network’s activity at the initial time step of each trial. We compared two learning paradigms:
Dynamic adaptation (left panel, \FigRef{figs5}.B): In this case, the contextual information—whether the food is to the left or right—is encoded in the neuronal activity, rather than in fixed synaptic weights. This allows the network to flexibly adjust its response from trial to trial without requiring long-term structural changes.
Plastic adaptation (right panel, \FigRef{figs5}.B): Here, learning occurs through synaptic weight updates. As a result, at the beginning of each trial, 
contextual information is encoded in the synaptic weights and not directly in the network activity. 
\paragraph{Implications for Neural Encoding and Experimental Validation}
At the start of each trial, the ant must initially move straight down the corridor before making a directional choice. Given this, it is not immediately obvious that any information about the correct target location should be present in the network’s state at the very beginning of the trial. Indeed, if contextual information were stored only in synaptic weights, it would not be explicitly represented in early network activity (as shown in Fig. B, right panel). However, when information about the subtask is dynamically encoded in neuronal activity, it becomes detectable through neural recordings (Fig. B, left panel). This finding suggests that adaptive behavior in biological systems may rely more on network dynamics and transient activity states, rather than purely on synaptic modifications.
Our results highlight a fundamental distinction between synaptic plasticity-based adaptation and dynamic adaptation via network activity, with important implications for understanding biological learning mechanisms and designing adaptive artificial systems.

\section*{Training the low-level network for Mujoco-ANT}

We conducted a reinforcement learning experiment using the MuJoCo-based \texttt{Ant-v4} environment in OpenAI Gym. The agent's objective was to learn movement patterns through a graded goal-directed reinforcement learning model.
The experiment was run for 500,000 episodes, with each episode lasting 400 timesteps.

The agent utilized a neural network model with 500 recurrent units, an input size of 31 (27 state variables + 4 target indicators, corresponding to UP, DOWN, LEFT, and RIGHT), and an output dimension of 8 (corresponding to the Ant's actuators). 

Each episode started with environment reset and the agent processing observations. The agent selected actions based on its internal policy, executed them, and received a reward. The reward was defined based on directional movement accuracy. Only readout weights where trained, following a policy gradient approach. Optimization was performed using the Adam optimizer.

\end{document}